\documentclass[numberedappendix]{emulateapj} % double-column, single spaced document
\usepackage{epsfig,color,pifont}
\usepackage{enumerate}
\usepackage{amsmath}

\usepackage{natbib}
\bibpunct{(}{)}{;}{a}{}{,}
\def\spose#1{\hbox to 0pt{#1\hss}}
\def\ltsimm{\mathrel{\spose{\lower 3pt\hbox{$\sim$}}
        \raise 2.0pt\hbox{$<$}}}
\def\gtsimm{\mathrel{\spose{\lower 3pt\hbox{$\sim$}}
        \raise 2.0pt\hbox{$>$}}}

\def\cm{{\rm\thinspace cm}}

\def\s{{\rm\thinspace s}}

\def\g{{\rm\thinspace g}}

\def\erg{{\rm\thinspace erg}}
\def\Hz{{\rm\thinspace Hz}}

\def\ster{{\rm\thinspace ster}}
\def\ergps{\hbox{${\rm\erg\s^{-1}\,}$}}

\def\pcm{\hbox{${\rm\cm^{-1}\,}$}}
\def\pcm2{\hbox{${\rm\cm^{-2}\,}$}}
\def\pcm3{\hbox{${\rm\cm^{-3}\,}$}}
\def\ergpscm3Hz{\hbox{${\rm\ergps\cm^{-3}\Hz^{-1}\,}$}}
\def\ergpscm3Hzster{\hbox{${\rm\ergps\cm^{-3}\Hz^{-1}\ster^{-1}\,}$}}
\def\gpcm3{\hbox{${\rm\g\cm^{-3}\,}$}}
\def\ergpcm2{\hbox{${\rm\erg\cm^{-2}\,}$}}
\def\ergpcm3{\hbox{${\rm\erg\cm^{-3}\,}$}}
\def\phpscm2{\hbox{${\rm photons\s^{-1}\cm^{-2}\,}$}}

\def\aap{{\rm A\&A}}
\def\apj{{\rm ApJ}}
\def\apjs{{\rm ApJS}}

\def\mnras{{\rm MNRAS}}

\def\jcp{{\rm J.~Comput.~Phys}}

\def\jetp{{\rm J.~Expl.~Theoret.~Phys.}}
\def\pnas{{\rm Proc.~Nat.~Acad.~Sci.}}
\begin{document} 

%\title{Exciting specific modes of the magnetorotational instability as
%  an efficient means of reaching steady state turbulence.}
\title{Equilibrium disks, MRI mode excitation, and steady state
  turbulence in global accretion disk simulations.}

\shorttitle{Exciting MRI modes in equilibrium disks}

\author{E.~R.~Parkin \& G.~V.~Bicknell}
\affil{Research School of
       Astronomy and Astrophysics, The Australian National University, Australia} 
\email{email: parkin@mso.anu.edu.au}

\shortauthors{E.~R.~Parkin \& G.~V.~Bicknell}

\label{firstpage}

\begin{abstract}
  Global three dimensional magnetohydrodynamic (MHD) simulations of
  turbulent accretion disks are presented which start from fully
  equilibrium initial conditions in which the magnetic forces are
  accounted for and the induction equation is satisfied. The local
  linear theory of the magnetorotational instability (MRI) is used as
  a predictor of the growth of magnetic field perturbations in the
  global simulations. The linear growth estimates and global
  simulations diverge when non-linear motions - perhaps triggered by
  the onset of turbulence - upset the velocity perturbations used to
  excite the MRI. The saturated state is found to be independent of
  the initially excited MRI mode, showing that once the disk has
  expelled the initially net flux field and settled into
  quasi-periodic oscillations in the toroidal magnetic flux, the
  dynamo cycle regulates the global saturation stress
  level. Furthermore, time-averaged measures of converged turbulence,
  such as the ratio of magnetic energies, are found to be in agreement
  with previous works. In particular, the globally averaged stress
  normalized to the gas pressure, $\overline{<\alpha_{\rm P}>} =
  0.034$, with notably higher values achieved for simulations with
  higher azimuthal resolution. Supplementary tests are performed using
  different numerical algorithms and resolutions. Convergence with
  resolution during the initial linear MRI growth phase is found for
  $23-35~$cells per scaleheight (in the vertical direction).
\end{abstract}

\keywords{accretion, accretion disks - magnetohydrodynamics -
  instabilities - turbulence}

\maketitle

\section{Introduction}
\label{sec:intro}

Accretion disks are ubiquitous in astrophysics and play an essential
part in the formation of stars and galaxies. For accretion through a
disk to be effective, angular momentum must be transported radially
outwards, allowing material to move radially inwards. One means of
achieving this is through viscous torques
\citep[][]{Lynden-Bell:1974}, and considerable progress has been made
using the phenomenological $\alpha$-viscosity introduced by
\cite{Shakura:1973} which assumes that viscosity is provided by
turbulent stresses which scale with the gas pressure. However, despite
its successes, the $\alpha$-viscosity model provides little physical
insight into the mechanism(s) responsible for the turbulent stress.

Even prior to the work of \cite{Shakura:1973}, instabilities in
magnetized rotating plasmas had been discovered by
\cite{Velikhov:1959} and \cite{Chandrasekhar:1960}. Yet, it was not
until the seminal work of \cite{BH91, BH98} that the so-called
magnetorotational instability (MRI) received widespread attention as
the agent responsible for the onset of accretion disk
turbulence. Linear stability analysis has shown that the MRI will
amplify a seed magnetic field indefinitely until confronted by the
strong-field limit or the diffusion scale \citep{BH92, Terquem:1996,
  Papaloizou:1997}. Non-linear stability analysis finds that growth of
the magnetic field by the linear phase of the MRI is likely to be
truncated by saturation resulting from secondary, or parasitic,
instabilities \citep[e.g.][]{Goodman:1994, Pessah:2010}. That
saturation of the magnetic field occurs was clearly demonstrated by
even the very first shearing box simulations \citep{Brandenburg:1995,
  Hawley:1995, Stone:1996}.

Contemplating the next steps in magnetized disks studies is aided by
summarising what we have already learned. For example, as mentioned
above, it is clear that the magnetic field reaches saturation and that
the resulting Maxwell stress dominates the angular momentum
transport. In numerical simulations this necessitates high resolution
to ensure that the fastest growing MRI modes are sufficiently well
resolved \citep[see, e.g.,][]{Sano:2004, Fromang:2006,
  Noble:2010,Flock:2011, Hawley:2011}. Related to this point is the
importance of stratification, which introduces a characteristic length
scale, removing the problem of non-convergence with simulation
resolution encountered in unstratified simulations
\citep{Fromang:2007a, Lesur:2007, Simon:2009, Guan:2009, Davis:2010,
  Sorathia:2012}. Stratification could also play a role in the dynamo
process which sets the saturation stress \citep{Brandenburg:2005,
  Vishniac:2009,Shi:2010, Gressel:2010}. However, the shearing box
approximation used in a large number of numerical studies to-date has
limitations \citep[e.g.][]{Regev:2008, Bodo:2008, Bodo:2011},
including the use of shearing-periodic boundary conditions in the
radial direction, and/or periodic boundary conditions in the vertical
direction. There boundary conditions artificially trap magnetic flux,
assisting the maintenance of the turbulent dynamo and obscuring the
dependence of the saturated state on resolution. This is supported by
a comparison of periodic and open boundary conditions in global models
by \cite{Fromang:2006} where the former were found to assist the
dynamo by preventing magnetic flux from being expelled from the
domain. In this regard global models have the advantage of removing
the unphysical influence of the shearing box boundary conditions,
albeit at a much larger computational expense.

Other motivations for global models are the results from stability
analyses of non-axisymmetric disturbances in magnetized accretion
disks, where the most robust MRI modes are localized and the most
robust buoyant (Parker) modes are global \citep{Terquem:1996,
  Papaloizou:1997}. Therefore, large radial extents are required to
accommodate the more global modes, and in this regard there is a limit
to the radial periodicity adopted in most shearing box
simulations. These factors point to the need for high resolution,
global, stratified disk simulations to further unravel the
complexities of magnetorotational turbulence. 

Of the global simulation studies that have been performed a large
number of the findings from local models have been maintained or have
persisted; the ratio of the Maxwell and Reynolds stress is $\sim3$,
and variations in toroidal magnetic field with time are suggestive of
a dynamo cycle \citep{Hawley:2000, Hawley:2002,
  Fromang:2006,Fromang:2009, Lyra:2008, Sorathia:2010, Sorathia:2012,
  Flock:2010, Flock:2011, Flock:2012, O'Neill:2011, Hawley:2011,
  Beckwith:2011, Mignone:2012, McKinney:2012,Romanova:2012}. However,
a large number of these simulations do not start from fully
equilibrium initial conditions where the magnetic field is accounted
for in the force balance and in the induction equation. Both local and
global models started with poloidal fields which do not satisfy the
induction equation show rapid disruption and re-arrangement of the
disk \citep[e.g.][]{Miller:2000, Hawley:2000, Hawley:2011}. This
introduces a transient phase where channel flows are fueled by rapid
shearing of the poloidal field lines. As such, extended run times are
required to ensure that transients have subsided. To our knowledge, no
previous global simulations of the MRI in stratified disks have used a
fully equilibrium initial disk (i.e. satisfying force balance {\it
  and} the induction equation).

We aim to explore the influence of magnetic fields on an accretion
disk with global simulations. In this first paper we present
equilibrium initial disk models with arbitrary radial density and
temperature profiles. We then investigate the saturation (both locally
and globally) of the growth of magnetic field perturbations. To this
end we excite the MRI in global simulations using linear MRI
calculations as a guide, and recover growth of magnetic field
perturbations in agreement with estimates. In so doing we show that
non-linear gas motions saturate the initial growth of the magnetic
field and that at later times the turbulent state retains no knowledge
of the initially excited MRI mode(s).

The plan of this paper is as follows: in \S~\ref{sec:model} we
describe the equilibrium initial conditions and details of the
numerical calculations and in \S~\ref{sec:excite} we perform a linear
perturbation analysis for the non-axisymmetric MRI. We present a suite
of global magnetized disk simulations in \S~\ref{sec:global_models},
which explore the effect of different MRI mode excitation and
numerical algorithms. In \S~\ref{sec:discussion} we compare our
results to previous work, and then close with conclusions in
\S~\ref{sec:conclusions}.
\pagebreak
\section{The model}
\label{sec:model}

\subsection{Simulation code}
\label{subsec:hydromodel}

For our global disk simulations, we use a 3D spherical
$(r,\theta,\phi)$ coordinate system with a domain which closely
encapsulates the initial disk \citep[e.g.][]{Fromang:2006}, and we
solve the time-dependent equations of ideal MHD using the
\textsc{PLUTO} code \citep{Mignone:2007}. We note that throughout this
work we describe our results in terms of both spherical
$(r,\theta,\phi)$ and/or cylindrical $(R,\phi,z)$ coordinates, with
$R=r \sin \theta$ and $z=r \cos \theta$. The relevant equations for
mass, momentum, energy conservation, and magnetic field induction are:
\begin{eqnarray}
\frac{\partial\rho}{\partial t} + \nabla \cdot [\rho {\bf v}] &  =  & 0, \\
\frac{\partial\rho{\bf v}}{\partial t} + \nabla\cdot[\rho{\bf vv} -
{\bf BB} + (P+\frac{1}{2}|{\bf B}|^2){\bf I}] & = & -\rho \nabla\Phi,\\
\frac{\partial E}{\partial t} + \nabla\cdot[(E + P){\bf v} - ({\bf v \cdot B}){\bf B}] &
=& -\rho {\bf v}\cdot \nabla\Phi -\rho \Lambda \nonumber\\\label{eqn:energy}\\
\frac{\partial\bf B}{\partial t} & = & \nabla \times ({\bf v \times
  B}) \label{eqn:induction}. 
\end{eqnarray}
\noindent Here $E = \rho\epsilon + \frac{1}{2}\rho|{\bf v}|^{2} +
\frac{1}{2}|{\bf B}|^2$, is the total gas energy density, $\epsilon$ is the
internal energy per unit mass, ${\bf v}$ is the gas velocity, $\rho$
is the mass density, and $P$ is the pressure. We use the ideal gas
equation of state, $\rho \epsilon = P/(\gamma - 1)$, where the
adiabatic index $\gamma = 5/3$. The adopted scalings for density,
velocity, temperature, and length are, respectively,
\begin{eqnarray}
  \rho_{\rm scale}&=&1.67\times10^{-7}~{\rm gm~s^{-1}}, \nonumber \\
  v_{0}&=&c, \nonumber \\
  T_{\rm scale}&=&\mu m c^{2}/k_{\rm B} = 6.5\times10^{12}~{\rm K},
  \nonumber \\
  l_{\rm scale}&=&1.48\times10^{13}~{\rm cm}, \nonumber 
\end{eqnarray}
\noindent where $c$ is the speed of light, and the value of $l_{\rm
  scale}$ corresponds to the gravitational radius of a $10^{8}~{\rm
  M_{\odot}}$ black hole. 

The gravitational potential, $\Phi$ of a central point mass (ignoring
self-gravity of the disk), $\Phi$ is modelled using the
pseudo-Newtonian potential introduced by \cite{Paczynsky:1980}:
\begin{equation}
\Phi = \frac{-1}{r - 2}. \label{eqn:pwpot}
\end{equation}
\noindent Note that we take the gravitational radius (in scaled
units), $r_{\rm g} = 1$. The Schwarzschild radius, $r_{\rm s}=2$ for a
spherical black hole and the innermost stable circular orbit (ISCO)
lies at $r=6$. The $\Lambda$ term on the RHS of Eq~(\ref{eqn:energy})
is an ad-hoc cooling term used to keep the scaleheight of the disk
approximately constant throughout the simulations; without any
explicit cooling in conjunction with an adiabatic equation of state,
dissipation of magnetic and kinetic energy leads to an increase in gas
pressure and, consequently, the disk scaleheight over time. The form
of $\Lambda$ is particularly simple,
\begin{equation}
  \Lambda = \frac{\rho}{(\gamma - 1)} \frac{T(R,z) -
    T_{\rm 0}(R)}{2 \pi R/v_{\phi}} \label{eqn:cooling_function}
\end{equation}
\noindent where $T_{0}(R)$ and $T(R,z)$ are the position dependent
initial and current temperature, respectively, $v_{\phi}$ is the
rotational velocity, and $R$ is the cylindrical radius. This cooling
function drives the temperature distribution in the disk back towards
the initial one over a timescale of an orbital period and is similar
in its purpose to the cooling functions used by \cite{Shafee:2008},
\cite{Noble:2010}, and \cite{O'Neill:2011}. Note that we only apply
cooling within $|z| < 2H$, where $H$ is the scaleheight of the disk,
allowing heating via dissipation to occur freely in the corona. Our
choice of an orbital period for the cooling timescale is somewhat
arbitrary but is chosen as it represents a characteristic timescale
for the disk.

The \textsc{PLUTO} code was configured to use the five-wave HLLD
Riemann solver of \cite{Miyoshi:2005}, piece-wise parabolic
reconstruction \citep[PPM -][]{Colella:1984}, and second-order
Runge-Kutta time-stepping. In order to maintain the ${\bf \nabla \cdot
  B} = 0$ constraint for the magnetic field we use the upwind
Constrained Transport (UCT) scheme of \cite{Gardiner:2008}. Such a
configuration has been shown to be effective in recovering the linear
growth rates of the axisymmetric MRI by \cite{Flock:2010}. In
\S~\ref{subsec:alg_test} we test a number of different numerical
setups: order of reconstruction, slope limiters, and simulation
resolution. However, in all of the other global simulations presented
in \S~\ref{sec:global_models} we use reconstruction on characteristic
variables \citep[e.g.][]{Rider:2007}. A Courant-Friedrichs-Lewy (CFL)
value of 0.35 was used for all simulations.

The grid used for the global simulations is uniform in the $r$ and
$\phi$ directions and extends from $r=4-34$ and $\phi = 0- \pi /2$. In
the $\theta$ direction we use a graded mesh which places slightly more
than half of the cells within $|z| \leq 2H$ of the disk mid-plane with
a uniform $\Delta \theta$, where $H$ is the thermal disk scaleheight,
and the remainder of the cells on a stretched mesh between $2H < |z| <
5H$. For example, for simulation gbl-$m10$ the 256 cells in the
$\theta$ direction are distributed so that 140 cells are uniformly
spaced between $2H < |z| < 5H$. The respective grid resolutions and
number of cells per scaleheight for the three global simulations are
noted in Table~\ref{tab:alg_test}. The grid cell aspect ratio at the
mid-plane of the disk and at a radius of $r=18.5$ (i.e. the disk
midpoint) are $ r\Delta \theta :\Delta r : r \Delta \phi = 1:1.4:8.6$
and $1:1.6:2.5$ for models gbl-$m10$ and gbl-$m10+$, respectively. The
$r$ and $\theta$ boundary conditions depend on whether the cell
adjacent to the boundary contains $>1\%$ disk material - which we
determine using a tracer variable. If this constraint is satisfied we
use outflow boundary conditions on all hydrodynamic variables except
$v_{\phi}$ which is determined from a zero-shear boundary condition
(i.e. $d \Omega / d r = 0$) and the normal velocity, for which we
enforce zero inflow. If the condition on disk material at the boundary
is not satisfied we use outflow boundary conditions on hydrodynamic
variables with the limit that the values must lie between the floor
values and the initial conditions for the background atmosphere - we
find this choice to be useful in setting up a steady background inflow
during the early stages of the simulation before material initially in
the disk evolves to fill the domain. For the magnetic field we use
zero gradient boundary conditions on the tangential field components
and allow the UCT algorithm to calculate the normal component so as to
satisfy the divergence free constraint, with the exception that at the
inner radial boundary we enforce a negative magnetic stress condition
\citep[e.g.][]{Stone:2001}. A periodic boundary condition is used in
the $\phi$ direction. Finally, we use floor density and pressure
values which scale linearly with radius and have values at the outer
radial boundary of $10^{-4}$ and $5\times10^{-9}$, respectively.

\subsection{Initial conditions}
\label{subsec:initialconditions}

Motivated by the fact that magnetorotationally turbulent disks are
dominated by toroidal field, we start from an analytic equilibrium
disk with a purely toroidal magnetic field. The disk equilibrium is
derived in axisymmetric cylindrical coordinates $(R,z)$; further
details can be found in Appendix~\ref{sec:appendix} along with
alternative disk solutions which may be of use in future work. In the
following we briefly summarise the equations for the isothermal in
height, $T = T(R)$, constant ratio of gas-to-magnetic pressure, $\beta
= 2P/|B|^2 \equiv 2P/B_{\phi}^2$, net magnetic flux disk adopted for
the simulations presented in this paper. The choice of temperature and
magnetic field lead to a density distribution, in scaled units,
\begin{equation}
  \rho(R,z) = \rho(R,0) \exp\left(\frac{-\{\Phi(R,z) - \Phi(R,0)\}}{T(R)} \frac{\beta}{1 + \beta} \right), \label{eqn:rho}
\end{equation}
\noindent where the pressure, $P = \rho T$. For the radial profiles
$\rho(R,0)$ and $T(R)$ we use simple functions inspired by the
\cite{Shakura:1973} disk model, except with an additional truncation
of the density profile at a specified outer radius:
\begin{eqnarray}
  \rho(R,0) & = & \rho_{0} f(R,R_{0},R_{\rm out})
  \left(\frac{R}{R_{0}}\right)^{\epsilon}, \\
  T (R) & = & T_{\rm 0} \left(\frac{R}{R_{0}}\right)^{\chi}, \label{eqn:temp}
\end{eqnarray}
\noindent where $\rho_{0}$ sets the density scale, $R_{0}$ and $R_{\rm
  out}$ are the radius of the inner and outer disk edge, respectively,
$f(R,R_{0},R_{\rm out})$ is a tapering function and is described in
Appendix~\ref{sec:appendix}, and $\epsilon$ and $\chi$ set the slope
of the density and temperature profiles, respectively. In all of the
global simulations $R_{0}=7$, $R_{\rm out}=30$, $\rho_{0}=10$,
$T_{0}=4.5\times10^{-4}$, $\epsilon=-33/20$, and $\chi=-9/10$
\citep[consistent with the radial scaling in the gas pressure and
Thomson-scattering opacity dominated region from][]{Shakura:1973},
producing disks with an aspect ratio, $H/R = 0.05$. The rotational
velocity of the disk is close to Keplerian, with a minor modification
due to the gas and magnetic pressure gradients,
\begin{eqnarray}
  v_{\phi}^2(R,z) = v_{\phi}^2(R,0) + \{\Phi(R,z) -
  \Phi(R,0)\}\frac{R}{T}\frac{d T}{d R}, \label{eqn:vphi}
\end{eqnarray}
\noindent where,
\begin{eqnarray}
v_{\phi}^2(R,0) =  R \frac{\partial
  \Phi (R,0)}{\partial R} + \frac{2 T}{\beta} + \nonumber \\
\left(\frac{1 + \beta}{\beta} \right)\left(\frac{R
    T}{\rho(R,0)}\frac{\partial \rho(R,0)}{\partial R}
 + R \frac{d T}{d R}\right).   \label{eqn:vphi0}
\end{eqnarray}
One advantage using such an equilibrium disk is that one begins with a
disk that is close to the expected scale height and density. An
isothermal disk, for example, has a scale height that is proportional
to $R^{3/2}$. 

Finally, the region outside of the disk is set to be an initially
stationary, spherically symmetric, hydrostatic atmosphere with a
temperature and density given by,
\begin{eqnarray}
  T_{\rm atm}(r) & = & -\frac{\Phi}{2}, \\
  \rho_{\rm atm} (r)  & = & \rho_{\rm atm} (r_{\rm
    ref}) \left(\frac{\Phi(r)} {\Phi(r_{\rm ref})} \right), 
\end{eqnarray}
\noindent where $\rho_{\rm atm}=4\times10^{-5}\rho_{0}$ and $r_{\rm
  ref}$ is a reference radius which we take to be $R_{\rm max}$, the
radius of peak disk density (see Appendix~\ref{sec:appendix}). The
transition between the disk and background atmosphere occurs where
their total pressures balance.

As an example, model gbl-$m10$ corresponds to a disk with a peak
density of $1.67\times10^{-7}~{\rm gm~s^{-1}}$ and a peak temperature
of $2.9\times10^{9}~{\rm K}$.

\subsection{Diagnostics}
\label{subsec:diagnostics}

Turbulence is by its very nature chaotic. Therefore, averaged
quantities are particularly useful diagnostics. In this section we
describe how we calculate averages, and define the variables used to
analyse the simulations.

To compute shell-averaged values (denoted by curly brackets) of a
variable $q$ at a radius $r$ we average in the $\theta$ and $\phi$
directions via,
\begin{equation}
 \{q\} = \frac{\int q r^2 \sin \theta d\theta d\phi}{\int r^2 \sin \theta d\theta d\phi}.
\end{equation}
\noindent Similarly, we calculate a horizontally averaged value
(denoted by square brackets) as,
\begin{equation}
  [q] = \frac{\int q r \sin \theta dr d\phi}{\int r \sin \theta dr d\phi}
\end{equation}
To attain a volume-averaged value (denoted by angled brackets) we
integrate over the radial profile of shell-averaged values and
normalize by the radial extent,
\begin{equation}
  <q> = \frac{\int \{q\} dr}{\int dr}
\end{equation}
\noindent Time averages receive an overbar, such that a volume and
time averaged quantity would read $\overline{<q>}$. (Note that
density-weighted averages are computed, but only for {\it
  hydrodynamical} variables.) For the analysis presented in
\S~\ref{sec:global_models} we restrict the integration over $r$ and
$\theta$ to the range $10<r<30$ and in $\pi/2 - \theta_{2H/R} <\theta<
\pi/2 + \theta_{2H/R}$, where $\theta_{2H/R}=\tan^{-1}(2H/R)$. We
define this region as the ``disk body'' and limit the integration over
this region to allow comparison against recent global
\citep[e.g.][]{Fromang:2006, Beckwith:2011,
  Sorathia:2010,Flock:2011,Flock:2012,Hawley:2011,Sorathia:2012} and
large local\footnote{\cite{Guan:2011} find that properties of
  turbulence within the disk body in stratified disks are
  quantitatively similar to those of unstratified disks \citep[see
  also][]{Hawley:1995,Stone:1996}.} simulations
\citep[e.g.][]{Guan:2011,Simon:2012}.

In order to keep a track of the fluctuations in the scaleheight of the
disk during the simulation - which results from the interplay between
adiabatic heating and our cooling function - a density-weighted
average disk scaleheight is computed, where we take $H/R = c_{\rm
  s}/v_{\phi}$ (where $c_{\rm s}$ is the sound speed), then perform a
density-weighted shell-average followed by a radial averaging to
acquire a volume averaged value, $<H/R>$.

For accretion to occur, angular momentum must be transported radially
outwards by turbulent stresses, and a major focus of numerical
simulations is quantifying the stress. To this end, we define the
perturbed flow velocity as\footnote{Using an azimuthally averaged
  velocity when calculating the perturbed velocity removes the
  influence of strong vertical and radial gradients
  \citep{Flock:2011}.}  $\delta v_{\rm i}=v_{\rm i}-\int v_{\rm i} r
\sin\theta d\phi / \int r \sin \theta d\phi$ with ${\rm i}={\rm
  R},\phi$, and compute the $R-\phi$ component of the combined
Reynolds and Maxwell stress,
\begin{equation}
  W_{R\phi} = \rho \delta v_{\rm R}\delta v_{\phi} - B_{\rm R}B_{\phi},
\end{equation}
\noindent which is normalized by the gas pressure to acquire,
\begin{equation}
  \{\alpha_{\rm P}\} = \frac{\{W_{R\phi}\}}{\{P\}}.
\end{equation}
\noindent Furthermore, we calculate the $R-\phi$ component of the
Maxwell stress normalized by the magnetic pressure,
\begin{equation}
  \{\alpha_{\rm M}\} = \frac{-2 \{B_{\rm R}B_{\phi}\}}{\{|B|^2\}}. \label{eqn:alpha_M}
\end{equation}

To examine the operation of dynamo activity in the disk we compute the
toroidal magnetic flux, which is defined as,
\begin{equation}
  \Psi_{\phi}(\phi) = \int \int B_{\phi}(\phi) r \sin\theta dr d\theta. \label{eqn:phi_flux}
\end{equation}

The ability of the simulations to resolve the fastest growing MRI
modes is quantified in the same fashion as \cite{Noble:2010} and
\cite{Hawley:2011}. The wavelength of the fastest growing MRI modes
with respect to the grid resolution in the $z$ and $\phi$ directions
are, respectively,
\begin{equation}
  Q_{\rm z}= \frac{\lambda_{\rm MRI-z}}{\Delta z} =
  \sqrt{\frac{16}{15}} \frac{2 \pi |v_
{A z}| r \sin \theta}{v_{\phi} \Delta z}, 
\end{equation}
\noindent and,
\begin{equation}
Q_{\phi}= \frac{\lambda_{\rm MRI-\phi}}{R \Delta \phi} = \frac{2 \pi |v_
{A \phi}|}{\Delta \phi},
\end{equation}
\noindent where $v_{\rm A z}$ and $v_{\rm A \phi}$ are the vertical
and azimuthal Alfv{\' e}n speeds, respectively, $\Delta \theta$ and
$\Delta \phi$ are the cell sizes in the $\theta$ and $\phi$
directions, respectively, and $\Delta z= \sqrt{(r \sin \theta \Delta
  \theta)^2 + (\Delta r \cos \theta)^2}$ is the corresponding cell
size in the $z$ direction. We define a single valued measure of
resolvability as the fraction of cells in the disk body ($|z|<2H$)
that have $Q>8$ \citep[e.g. ][]{Sorathia:2012},
\begin{equation}
  N_{\rm i} = \frac{\Sigma C(Q_{\rm i}>8)}{\Sigma C} \label{eqn:resolvability}
\end{equation}
\noindent where ${\rm i}=z,\phi$ and $C$ represents a cell. The
principal aim of calculating $N_{\rm z}$ and $N_{\phi}$ is to quantify
how well resolved the turbulent state is in a simulation, and
consequently whether global simulations are approaching the region of
convergence found from shearing box simulations
\citep[][]{Hawley:2011}.

\subsection{Fourier analysis}
\label{subsec:fourier}

To allow a direct comparison between the growth of MRI modes estimated
from a linear perturbation analysis (\S~\ref{sec:excite}) and the
results of global simulations (\S~\ref{sec:global_models}) we analyse
the growth of magnetic field perturbations in Fourier space. The
procedure we follow is to remap the disk body (which we define in
\S~\ref{subsec:diagnostics}) to a cylindrical mesh with uniform cell
spacing in all directions, and a sufficiently fine resolution to
ensure that the smallest cells from the spherical simulation grid are
sampled. We then perform a 3D Fourier Transform of the data on the
cylindrical grid. A detailed description of the cylindrical Fourier
transform can be found in Appendix~\ref{sec:cft}.

\section{Exciting the MRI}
\label{sec:excite}

Given that our global simulations commence with an equilibrium disk
the MRI requires a seed perturbation to excite the growth of the
magnetic field and development of turbulence. For this purpose we have
chosen to excite a specific Fourier mode of the MRI using poloidal
velocity perturbations. In the following we present perturbation
calculations for the local, linear, non-axisymmetric MRI, the results
of which are used in \S~\ref{sec:global_models} to elucidate the
evolution of magnetic field perturbations in global numerical
simulations.

\begin{table}
\begin{center}
  \caption[]{Parameters used for the linear MRI growth
    calculations} \label{tab:local_models}
\begin{tabular}{lllllll}
%\begin{small}
\hline
Model & $\beta$ & $m$ & $k_{\rm z}$ & $k_{\rm R}$ & $\left(\frac{{\bf
      k \cdot v_{\rm A}}}{\Omega}\right)^2$ & $\delta v_{0}/c_{\rm s}$\\
\hline
lin-$m10$ & 20 & 10 & 5 & 2.5 & 0.03 & 0.1 \\
lin-$m10$-$\beta$100 & 100 & 10 & 5 & 2.5 & 0.006 & 0.1 \\
lin-$m10$-$\beta$300 & 300 & 10 & 5 & 2.5 & 0.002 & 0.1 \\
lin-$m10$-$\beta 300s$ & 300 & 10 & 5 & 2.5 & 0.002 & 0.3 \\
lin-$m40$ & 20 & 40 & 80 & 40 & 0.45 & 0.1 \\
\hline
%\end{small}
\end{tabular}
%\tablecomments{}
\end{center}
\end{table}

\subsection{Linear MRI growth models}

Studies of the linear, non-axisymmetric MRI in weakly magnetized disks
have been examined by a number of authors \citep{BH92, Terquem:1996,
  Papaloizou:1997}. \cite{BH92}'s local study showed that even if the
seed magnetic field is purely toroidal then the instability is still
present, albeit with growth rates roughly an order of magnitude lower
than those found for initially poloidal fields \citep{BH91}. This
result was supported by growth timescales approaching an orbital
period (for certain parameters) in more-global calculations by
\cite{Terquem:1996} where radial gradients were
preserved. Furthermore, these authors found that in the $k_{\rm
  z}/k_{\rm R} \ll 1$ limit - the primary domain of the MRI -
instabilities become increasingly localized with time. On the other
hand, in the $k_{\rm R}/k_{\rm z} \ll 1$ limit the Parker instability
dominates. In fact, even in the presence of dissipation, MRI modes
continue to become increasingly localized over time due to the time
dependence of the radial wavenumber \citep{Papaloizou:1997}. Common to
these studies is the finding that the non-axisymmetric MRI acts as a
mechanism for the transient amplification of seed magnetic/velocity
field perturbations by many orders of magnitude over tens of
orbits. One question is, how well does this immense field
amplification carry through to global, fully non-linear simulations?
To answer this one needs an estimate of the linear growth. In this
regard our analysis of the {\it non-axisymmetric} MRI in this paper is
complementary to studies of the {\it axisymmetric} MRI in previous
simulations \citep[e.g.][]{Hawley:1991, Flock:2010}.

To construct our prediction for the global simulations we utilize the
linear MRI model of \cite{BH92}. (The perturbation analysis used to
quantify the linear MRI growth is performed in cylindrical coordinates
($R,\phi,z$), whereas the global models presented in
\S~\ref{sec:global_models} are performed in spherical coordinates
($r,\theta,\phi$).) In brief, \citeauthor{BH92} perform a linear
stability analysis of a local patch of a disk using the shearing-sheet
approximation \citep{Goldreich:1965} where the perturbations are
assumed to have a spatial dependence $\exp[i(k_{\rm R}R + m \phi +
k_{\rm z}z )]$. The equations for the evolution of the magnetic field
perturbations form a pair of coupled second-order ordinary
differential equations\footnote{Note that there is a typographical
  error in equation (2.19) of \cite{BH92} where the final term should
  read $\delta B_{\rm z} N^2 (k_{\rm z}^2 - k^2)/k^2$.}. We let $N$ be
the Brunt-V\"{a}is\"{a}l\"{a} frequency, which for the equilibrium
disk described in \S~\ref{subsec:initialconditions} is,
\begin{equation}
  N^2 = \frac{2}{5}\frac{1}{T}\left(\frac{\beta}{1 +
      \beta}\right)^2 \left(\frac{z}{R}\right)^2 \frac{1}{(r-2)^4}, \label{eqn:N2}
\end{equation}
and define an independent time variable,
\begin{eqnarray}
  \tau & = & k_{\rm R}(t) R = k_{\rm R}(t=0) R - m \frac{d \Omega}{d \ln R}t,
  \label{eqn:tau}\\
  k^2 & = & k^2_{\rm R} + \frac{m^2}{R^2} + k^2_{\rm z}.
\end{eqnarray}
Replacing the angular velocity with that due to a Paczynski-Wiita
potential in the thin disk limit (i.e. $H/R \ll 1$), $\Omega^2 =
1/R^2(R -2)$, the equations describing linear perturbations
are\footnote{The angular velocity resulting from our disk model (cf
  Eqs~(\ref{eqn:vphi}) and (\ref{eqn:vphi0})) actually includes a
  small offset to Keplerian rotation. However, we find that this makes
  little difference to the perturbation calculations, and the
  subsequent comparison against global simulations in
  \S~\ref{sec:global_models}. Therefore, for the sake of simplicity,
  we adopt a purely Keplerian rotation profile for the local
  calculations.},
\begin{eqnarray}
\frac{d^2 \delta B_{\rm z}}{d \tau^2} = \frac{2 k_{\rm z}}{R k^2 (3R
  -2)}\left(\frac{2 \tau^2}{m^2}(R-2) - R -2 \right)\frac{d \delta
  B_{\rm R}}{d \tau} \nonumber\\
- \frac{4}{m^2}\left(\frac{R - 2}{3R - 2} \right)^2 \delta B_{\rm z}
\left[ \frac{({\bf k \cdot v_{\rm A}})^2}{\Omega^2}
  + \left( \frac{k^2 - k_{\rm z}^2}{k^2} \right)
  \frac{N^2}{\Omega^2}\right] \nonumber\\
+ \frac{4 k_{\rm z} \tau}{k^2
  m^2} \left( \frac{R - 2}{3R - 2} \right) \tau \frac{d \delta B_{\rm
    z}}{d \tau}, \nonumber \\ \label{eqn:Bz_pert}
\end{eqnarray}
\begin{eqnarray}
\frac{d^2 \delta B_{\rm R}}{d \tau^2} = -\frac{4}{m^2}\left(1 - \frac{k_{\rm
  R}^2}{k^2}\right)\frac{R - 2}{3R - 2}Rk_{\rm z}\frac{d \delta B_{\rm
z}}{d \tau} \nonumber\\
+ \frac{2}{R^2 k^2}\frac{R - 2}{3R - 2} \left[ \frac{2}{m^2}(\tau^2 - R^2k^2) +
  \frac{R+2}{R-2}\right]\tau\frac{d \delta B_{\rm R}}{d \tau} \nonumber \\
- \frac{4}{m^2} \left(\frac{R - 2}{3R - 2}\right)^2 \left[ \frac{({\bf k
    \cdot v_{\rm A}})^2}{\Omega^2} \delta B_{\rm R} - \frac{k_{\rm
    z}}{R k^2}\frac{N^2}{\Omega^2} \tau \delta B_{\rm z} \right]
\nonumber \\ \label{eqn:BR_pert}
\end{eqnarray}
\noindent where $\delta B_{\rm z}$ and $\delta B_{\rm R}$ are the
vertical and radial magnetic field perturbations. 

The time dependence of $k_{\rm R}$ in Eq~(\ref{eqn:tau}) is a
consequence of the radial wavenumber being sheared. Therefore, within
the framework of the \cite{BH92} analysis the radial wavenumber can
grow indefinitely so that radial disturbances can evolve to
arbitrarily small spatial extent. Clearly, when we come to making a
comparison against our global simulations, this will not be the case
due to finite numerical resolution.

The magnetic field perturbations are related through the
divergence-free constraint,
\begin{equation}
  {\bf k \cdot \delta B} = k_{\rm R} \delta B_{\rm R} + \frac{m}{R}\delta B_{\phi} +
  k_{\rm z}\delta B_{\rm z} = 0. \label{eqn:kdotdB}
\end{equation}

The unperturbed magnetic field topology only enters through ${\bf k
  \cdot v_{\rm A}}$. For our initially purely toroidal magnetic field
one finds,
\begin{equation}
\left(\frac{{\bf k \cdot v_{\rm A}}}{\Omega}\right)^2 = \frac{2  (R - 2) m^2 T
 }{\beta}, \label{eqn:kdotva2}
\end{equation}
\noindent To initiate the MRI we use the $R$ and $z$ components of the
linearized induction equation,
\begin{equation}
  \frac{d \delta B_{\rm R}}{dt} = i ({\bf k \cdot B})\delta v_{\rm R},  \label{eqn:dBRdt}  
\end{equation} 
\noindent and,
\begin{equation}
  \frac{d \delta B_{\rm z}}{dt} = i ({\bf k \cdot B})\delta v_{\rm z},\label{eqn:dBzdt}  
\end{equation} 
\noindent where $\delta v_{\rm R}$ and $\delta v_{\rm z}$ are the
poloidal velocity perturbations (with the imaginary part of $\delta v$
corresponding to the real part of $d \delta B / dt$). For the
perturbations in the $z$-components in both the linear MRI and global
calculations we use a waveform,
\begin{equation}
  \delta v_{\rm z} = \delta v_{\rm 0} \cos(k_{\rm R}R + m \phi +
  k_{\rm z}z), \label {eqn:vz_pert}
\end{equation}
\noindent which, on substitution into Eq~(\ref{eqn:dBzdt}), and with
the conversion between real and imaginary parts accounted for by a
phase shift in the trigonometric term, leads to,
\begin{equation}
  \frac{d \delta B_{\rm z}}{dt} = \delta v_{\rm
    0}\frac{2B_{\phi}}{v_{\phi}}\frac{R - 2}{3R - 2}\sin(k_{\rm R}R +
  m \phi + k_{\rm z}z), \label{eqn:dBz_pert}
\end{equation}
\noindent where $\delta v_{\rm 0}$ is the amplitude of the initial
velocity perturbations. An equivalent treatment to
Eq~(\ref{eqn:dBz_pert}) is used for the perturbations in the
$R$-components with the difference that we make use of the
incompressibility condition, ${\bf k \cdot \delta v}=0$, and set,
\begin{equation}
  \delta v_{\rm R} = \delta v_{\rm 0} \frac{k_{\rm z}}{k_{\rm R}} \cos(k_{\rm R}R + m \phi +
  k_{\rm z}z), \label {eqn:vR_pert}
\end{equation}
\noindent The remaining parameters used in the calculations are
summarised in Table~\ref{tab:local_models}.

\begin{figure}
  \begin{center}
    \begin{tabular}{c}
\resizebox{75mm}{!}{\includegraphics{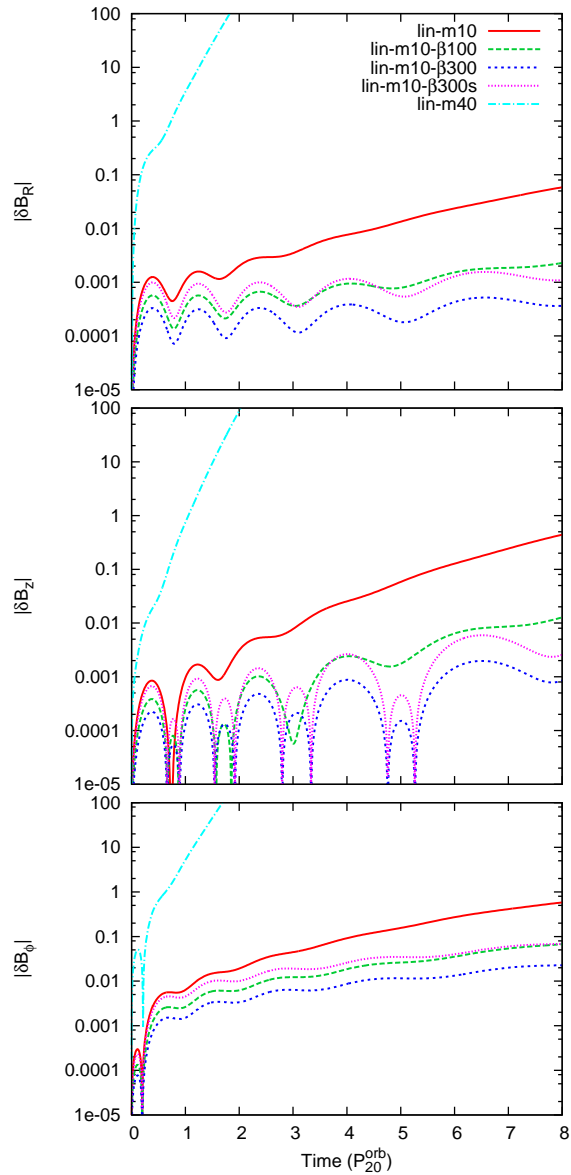}} \\
    \end{tabular}
    \caption{The evolution of magnetic field perturbations from linear
      MRI growth calculations showing successful magnetic field
      amplification. From top to bottom: $\delta B_{\rm R}, \delta
      B_{\rm z}$, and $\delta B_{\phi}$. The parameters used in these
      calculations are provided in Table~\ref{tab:local_models}.}
    \label{fig:std_local_models}
  \end{center}
\end{figure}

\begin{figure}
  \begin{center}
    \begin{tabular}{c}
\resizebox{75mm}{!}{\includegraphics{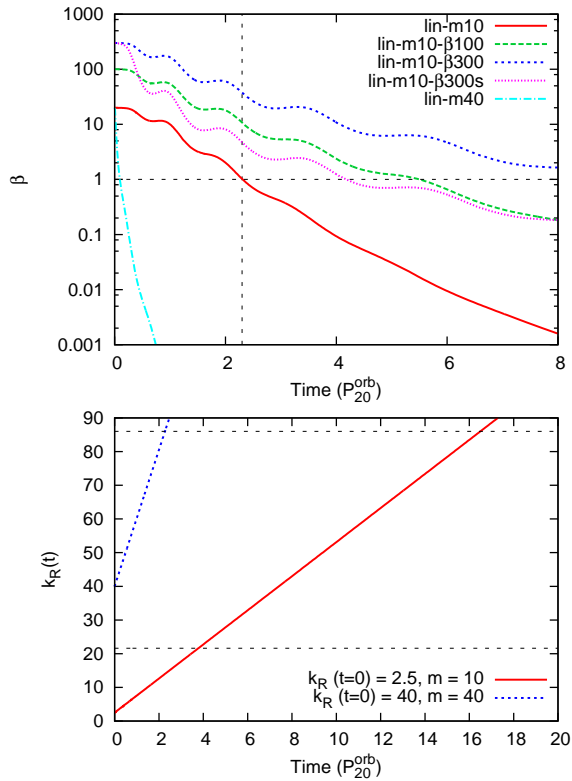}} \\
  \end{tabular}
  \caption{(Top): The plasma-$\beta$ calculated from the linear MRI
    growth calculations - see Table~\ref{tab:local_models} for the
    list of models and parameter values pertaining to each
    calculation. The vertical line indicates when model lin-$m10$
    reaches $\beta=1$. (Bottom): The time-dependent radial
    wavenumber. The red and blue lines corresponds to models lin-$m10$
    and lin-$m40$, respectively. The horizontal lines indicate the
    $k_{\rm R}$ values at the Nyquist limit for the global simulations
    gbl-$m10$ and gbl-$m10$-lllr (see Tables~\ref{tab:global_models}
    and \ref{tab:alg_test} and \S~\ref{sec:global_models}).}
    \label{fig:local_limits}
  \end{center}
\end{figure}

Our first calculation, model lin-$m10$, uses a $\beta=20$ magnetic
field and wavenumbers for the excited MRI mode of $m=10$, $k_{\rm z} =
5$, and $k_{\rm R}=2.5$. These wavenumbers are chosen to ensure
sufficient resolution in the global simulations and we leave a more
detailed discussion to \S~\ref{sec:global_models}. The amplitude of
the initial velocity perturbations, $\delta v_{0}$, is set to
$0.1~c_{\rm s}$, where $c_{\rm s}$ ($=\sqrt{T}$) is the sound
speed. Since we intend to use these calculations as a guide for our
global simulations, we use the equilibrium disk model described in
\S~\ref{subsec:initialconditions} to choose the input density and
temperature. Calculations are performed at a cylindrical radius,
$R=20$, and at the disk mid-plane where $N^2 = 0$ (see
Eq~(\ref{eqn:N2})). From Eq~(\ref{eqn:temp}) the disk temperature, $T
= 1.75\times10^{-4}$, and the density, $\rho=0.46$. The initial
components of $\delta{\bf B}$ are set to zero, so too is the initial
azimuthal velocity perturbation, $\delta v_{\phi}$ - the poloidal
velocity perturbations seed the instability through the $d\delta B/dt$
terms. To integrate Eqs~(\ref{eqn:Bz_pert}) and (\ref{eqn:BR_pert}) we
use an adaptive stepsize, 4th-order, explicit Runge-Kutta method
\citep[][]{Press:1986}. As Fig.~\ref{fig:std_local_models} shows, the
magnetic field perturbations grow extremely rapidly over the first few
$P^{\rm orb}_{\rm 20}$ with noticeable oscillations, where $P^{\rm
  orb}_{\rm j}$ is the radially dependent orbital period of the disk
at cylindrical radius j. The upper panel of
Fig.~\ref{fig:local_limits} shows the effective $\beta$ for the MRI
mode - the time required for the magnetic field to grow to $\beta=1$
is only a few orbital periods for model lin-$m10$. Evaluating the
approximate growth rate, $\omega$ of the magnetic energy, $\beta^{-1}$
(as the gas pressure remains constant) via $\beta^{-1}=\beta^{-1}_0
\exp(\omega t)$, we find an average growth rate over the first six
orbits, $\overline{\omega} = 0.14~\Omega$. Applying the same approach
to $\delta B_{\rm R}$ we find $\overline{\omega}=0.09\Omega$. This is
consistent with the findings of \cite{Terquem:1996} but is roughly an
order of magnitude larger than values of a few percent of the orbital
frequency quoted {\it in general} for the development of the
non-axisymmetric MRI by \cite{BH92}. Keeping all parameters fixed and
then varying the initial magnetic field strength, one sees from models
lin-$m10$-$\beta$100 and lin-$m10$-$\beta$300 the trend that the
growth rate of $\delta {\bf B}$ decreases with increasing initial
$\beta$. In model lin-$m10$-$\beta 300s$ the size of the initial
velocity perturbations is increased to $\delta v_{0} = 0.3~c_{\rm s}$
with the result that over the very first few orbits the growth of
$\delta B$'s becomes very similar to that of a stronger initial field
strength excited by smaller velocity perturbations. For a higher
wavenumber perturbation the rate of initial growth increases, as
evidenced by model lin-$m40$ (see Figs.~\ref{fig:std_local_models} and
\ref{fig:local_limits}). Evaluating the approximate growth rate of the
magnetic field energy and $\delta B_{\rm R}$ for model lin-$m40$
gives, $\overline{\omega}= 0.68~\Omega$ and $0.25\Omega$,
respectively\footnote{For comparison, the maximum growth rate for the
  {\it axisymmetric} MRI is $0.75\Omega$ \citep{BH91}.}. From these
results one may predict that the development of $\delta {\bf B}$ in
simulations will depend on the initial field strength and/or the
wavenumber of the excited mode(s). In \S~\ref{subsec:trigger} we
examine if this result holds true in global simulations.

\begin{figure}
  \begin{center}
    \begin{tabular}{c}
\resizebox{75mm}{!}{\includegraphics{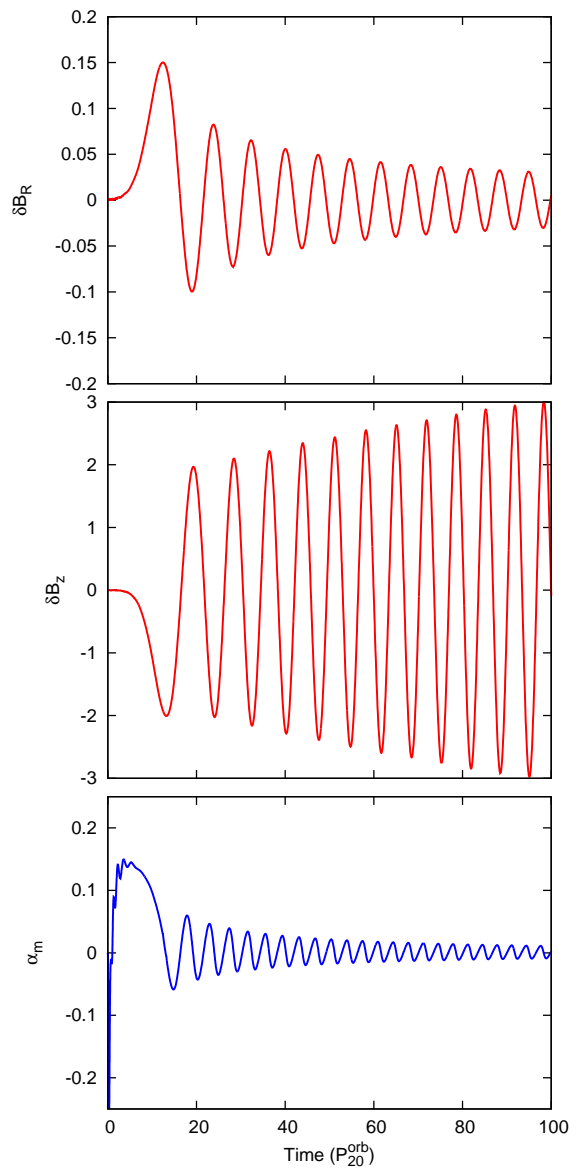}} \\
    \end{tabular}
    \caption{Evolution of $\delta B_{\rm R}$ (upper), $\delta B_{\rm
        z}$ (middle), and $\alpha_{\rm m}$ (lower) as a function of
      time in model lin-$m10$ over a longer time duration than shown
      in Fig.~\ref{fig:std_local_models}.}
    \label{fig:damp_local_models}
  \end{center}
\end{figure}

\cite{BH92} discuss the parameter $({\bf k \cdot v_{\rm
    A}})^2/\Omega^2$ and attribute to it an important role in the
ability of the MRI to successfully amplify the seed field. They find
that for $({\bf k \cdot v_{\rm A}})^2/\Omega^2 \gtsimm 2.9$ the
instability is stabilized and magnetic field oscillations are
damped. For the models shown in Fig.~\ref{fig:std_local_models}, this
parameter is much less than unity. From Eq~(\ref{eqn:kdotva2}) for
$({\bf k \cdot v_{\rm A}})^2/\Omega^2$ one can see that to increase
the value of this variable one can either decrease $\beta$ - which
increases the tension along field lines - or employ higher azimuthal
wavenumbers, $m$. The latter has the side-effect of increasing the
growth rate of $k_{\rm R}$ and causing tight wave crest wrapping, both
of which lead to a more rapid stabilization of the radial
disturbances. However, we find that irrespective of the value of
$({\bf k \cdot v_{\rm A}})^2/\Omega^2$, which is $0.03$ for lin-$m10$
(Table~\ref{tab:local_models}), the perturbation in the magnetic field
ultimately decays. This is shown in Fig~\ref{fig:damp_local_models}
(upper and middle panels) where the lin-$m10$ calculation is plotted
for a longer time duration. Despite continuing growth in $\delta
B_{\rm z}$, there is decay in $\delta B_{\rm R}$, which is a
consequence of the increase of $k_{\rm R}(t)$ combined with the
divergence-free constraint (Eq~(\ref{eqn:kdotdB})). The ratio of the
Maxwell stress to magnetic pressure ($\alpha_{\rm m}$) predicted from
the linear MRI growth calculations is shown in the lower panel of
Fig.~\ref{fig:damp_local_models}, where we define,
\begin{equation}
\alpha_{\rm m} = \frac{2 \delta B_{\rm R} (\delta B_{\phi} + B_{\phi})}{B^2}, B^2 = B_{\phi}^2 + \delta B_{\phi}^2 + \delta B_{\rm R}^2 + \delta
  B_{\rm z}^2.
\end{equation}
\noindent Clearly, considering that $\alpha_{\rm m}$, or to be more
exact\goodbreak$<\alpha_{\rm M}>$ (its global analogue - see
Eq~(\ref{eqn:alpha_M})), is a commonly used diagnostic in numerical
simulations \citep{Hawley:1995}. Under the action of the linear MRI
alone $<\alpha_{\rm M}>$ would never reach a steady value. This
ultimate decay of linear MRI disturbances is consistent with
\cite{Terquem:1996}'s finding of transient instability growth in a
number of numerical tests, in which $k_{\rm z} > k_{\rm R}$
initially. Shear causes $k_{\rm R}$ to grow but once $k_{\rm R} >
k_{\rm z}$ growth halts.

The ultimate decay of linear MRI modes has significance for global
models because the maintenance of dynamo action requires all
components of the field to be sufficiently strong. This highlights the
need for an additional mechanism, other than the linear MRI growth, to
replenish $\delta B_{\rm R}$ (e.g. parasitic instabilities -
\citeauthor{Goodman:1994}~\citeyear{Goodman:1994}; Parker instability
- \citeauthor{Tout:1992}~\citeyear{Tout:1992},
\citeauthor{Vishniac:2009}~\citeyear{Vishniac:2009}; dynamo action in
the steady-state turbulence
\citeauthor{Brandenburg:1995}~\citeyear{Brandenburg:1995},
\citeauthor{Hawley:1996}~\citeyear{Hawley:1996}).

The linear MRI growth calculations act as a check on our global
simulations, principally to examine whether our setup recovers the
growth rates of the linear MRI accurately. However, there is a limit
to the time interval when we can confidently make a comparison between
the linear growth models and global simulations. Firstly, the analysis
of \cite{BH92} adopts the Boussinesq approximation which becomes
invalid when the azimuthal magnetic field becomes super-thermal. The
upper panel of Fig.~\ref{fig:local_limits} shows that this limit is
reached in approximately 2.3 orbits for lin-$m10$ and 0.1 orbits for
lin-$m40$. Secondly, $k_{\rm R}(t)$ can grow indefinitely in the
linear growth models, yet this is not the case for our global
simulations which are restricted by finite numerical
resolution. Taking the Nyquist limit to be 2 grid cells, and
considering, for example, the resolution of model gbl-$m10$, the
maximum resolvable radial wavenumber is $k_{\rm R-Nyquist}=86$. This
limit is reached after $\sim 16.5$ and 2.3 orbits for models lin-$m10$
and lin-$40$, respectively (lower panel of
Fig.~\ref{fig:local_limits}). Therefore, choosing to excite a higher
wavenumber MRI mode limits the time interval where comparisons can be
made against linear perturbation theory, and this is one reason why we
choose to excite a lower wavenumber mode ($m=10$) in the global
simulations.

\begin{table*}
\begin{center}
  \caption[]{Global simulations and the corresponding linear growth rate.} \label{tab:alg_test}
\begin{tabular}{llllllll}
%\begin{small}
\hline
Model & Resolution & Reconstruction & Limiter & $n_{\rm r}/H$ &
$n_{\theta}/H$ & $n_{\phi}/H$ & $\omega_{\rm approx}$ \\
          & ($n_{\rm r}\times n_{\theta} \times n_{\phi}$) & & &
          ($10<r<30$) & ($|z|<2H$) & & ($\Omega(R=20)$) \\
\hline
lin-$m10$ & $-$ & $-$ & $-$ & $-$ & $-$ & $-$ & 0.09 $\pm$ 0.01 \\
gbl-$m10$ & $768 \times 256 \times 128$ & Parabolic & Char & 18-77 &
35 & 4 & 0.15 $\pm$ 0.03 \\
gbl-$m10+$ & $512 \times 170 \times 320$ & Parabolic & Char & 12-51 &
27 & 12.5 & 0.11 $\pm$ 0.02 \\
gbl-rand & $512 \times 170 \times 320$ & Parabolic & Char & 12-51 & 27
& 12.5 & 0.15 $\pm$ 0.04  \\
gbl-$m10$-lin & $768 \times 256 \times 128$ & Linear & Char & 18-77
&35 & 4 & 0.16 $\pm$ 0.03 \\
gbl-$m10$-cw   & $768 \times 256 \times 128$ & Parabolic & CW84 &
18-77 &35 & 4 & 0.13 $\pm$ 0.02 \\
gbl-$m10$-cs   & $768 \times 256 \times 128$ & Parabolic & CS08 &
18-77 &35 & 4 & 0.14 $\pm$ 0.02 \\
gbl-$m10$-hr   & $896 \times 300 \times 150$ & Parabolic & Char &
21-90 & 41 & 5 & 0.16 $\pm$ 0.04 \\
gbl-$m10$-lr   & $512 \times 170 \times 96$ & Parabolic & Char & 12-51
& 23 & 3 & 0.13 $\pm$ 0.03 \\
gbl-$m10$-llr   & $342 \times 112 \times 64$ & Parabolic & Char & 8-34
& 15 & 2 & 0.09 $\pm$ 0.04 \\
gbl-$m10$-lllr   & $192 \times 64 \times 32$ & Parabolic & Char & 5-19
& 9 & 1 & 0.05 $\pm$ 0.03 \\
\hline
%\end{small}
\end{tabular}
%\tablecomments{}
\end{center}
\end{table*}

\begin{table*}
\begin{center}
\caption[]{Time averaged quantities from the global simulations..} \label{tab:global_models}
\begin{tabular}{llllllllllllll}
%\begin{small}
  \hline
  Model & $m$ & $k_{\rm z}$ & $k_{\rm R}$ &
  $\frac{\delta v_{0}}{c_{\rm s}}$ & $\delta t_{\rm av}$$^{\rm a}$ & $\overline{N_{\rm z}}$ &
  $\overline{N_{\phi}}$
  &$\overline{\frac{<B_{\rm R}^2>}{<B_{\phi}^2>}}$ & $\overline{\frac{<B_{\rm
        z}^2>}{<B_{\rm R}^2>}}$ & $\overline{\frac{<B_{\rm z}^2>}{<B_{\phi}^2>}}$ &
  $\overline{<\alpha_{\rm P}>}$ & $\overline{<\alpha_{\rm M}>}$ &
  $\overline{<\beta>}$$^{\rm b}$ \\
  \hline
  gbl-$m10$ & 10 & 5 & 2.5 & 0.1 & 10-28 & 0.32 & 0.33 & 0.071
  & 0.25 & 0.018 & 0.017 & 0.31 & 32 \\
  gbl-$m10+$ & 10 & 5 & 2.5 & 0.001 & 12-26 & 0.45 & 0.75 & 0.12 & 0.29 & 0.036 & 0.034 & 0.41 & 17 \\
  gbl-rand & $-$ & $-$ & $-$ & 0.001 & 16-26 & 0.45 & 0.75 & 0.13 &
  0.30 & 0.037 & 0.034 & 0.41 & 17 \\
  \hline
%\end{small}
\end{tabular}
\tablecomments{$^{\rm a}$ Time interval over which averaging was performed,
  \newline
                          $^{\rm b}$ Time averaged plasma-$\beta$ in the disk body.}
\end{center}
\end{table*}

\begin{figure}
  \begin{center}
    \begin{tabular}{c}
\resizebox{75mm}{!}{\includegraphics{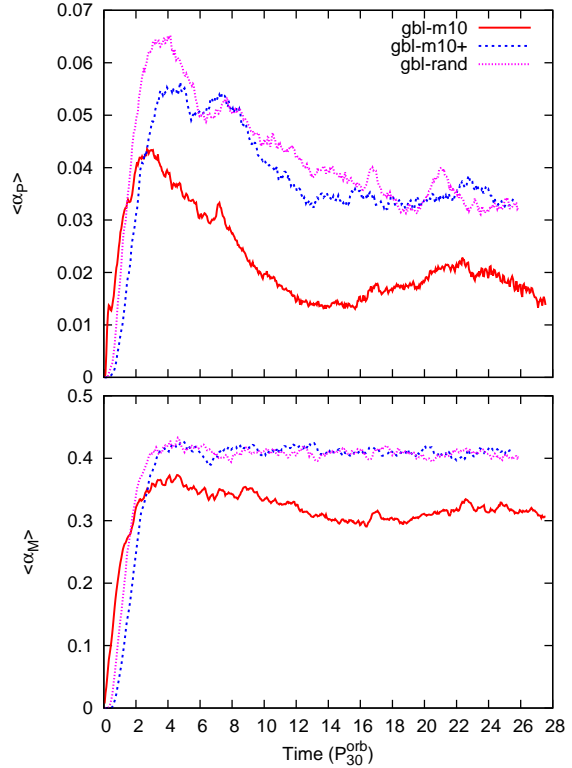}} \\
    \end{tabular}
    \caption{The evolution of $<\alpha_{\rm P}>$ (upper) and
      $<\alpha_{\rm M}>$ (lower) in the global models as a function of
      time in units of the orbital period at $R=30$. See
      Table~\ref{tab:global_models} for a list of the models and time
      averaged values.}
    \label{fig:global_stress}
  \end{center}
\end{figure}

\begin{figure}
  \begin{center}
    \begin{tabular}{c}
    \resizebox{75mm}{!}{\includegraphics{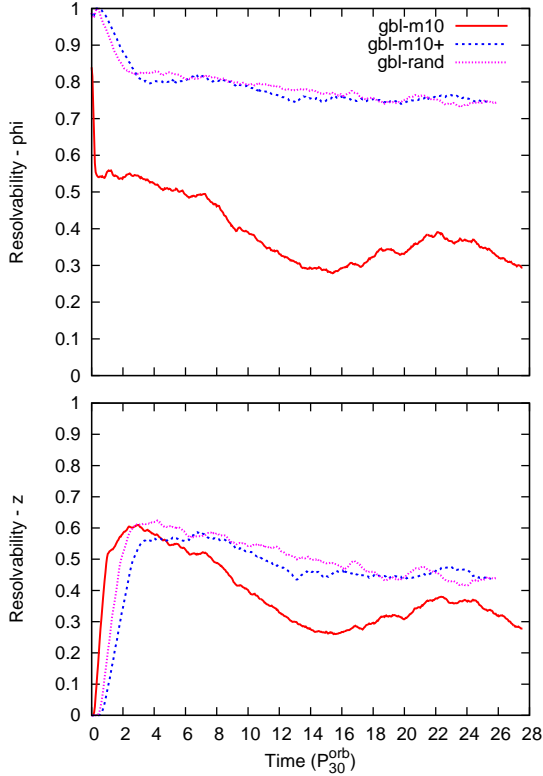}} \\
    \end{tabular}
    \caption{Resolvability of the MRI in the $\phi$ direction (upper)
      and $z$ direction (lower) in models gbl-$m10$, gbl-$m10+$, and
      gbl-rand. For a definition of the resolvability see
      Eq~(\ref{eqn:resolvability}) and \S~\ref{subsec:diagnostics}.}
    \label{fig:resolvability}
  \end{center}
\end{figure}

\begin{figure}
  \begin{center}
    \begin{tabular}{c}
\resizebox{75mm}{!}{\includegraphics{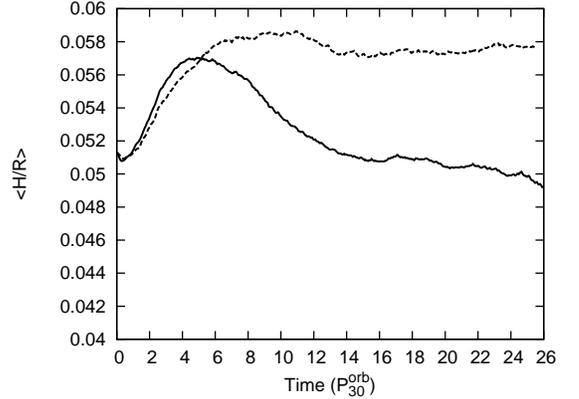}} \\
   \end{tabular}
   \caption{Density-weighted and volume averaged scaleheight of the
     disk, $<H/R>$ as a function of time in models gbl-$m10$ (solid
     line) and gbl-$m10+$ (dashed line).}
    \label{fig:H_R}
  \end{center}
\end{figure}
\pagebreak
\section{Global models}
\label{sec:global_models}

In this section we describe the results of global simulations using
the initial conditions and simulation setup described in
\S~\ref{sec:model}. The global simulations are listed along with grid
dimensions, number of cells per scaleheight, and approximate MRI
growth rates in \ref{tab:alg_test}. Time and volume averaged variables
quantifying the steady-state turbulence are given in
Table~\ref{tab:global_models}. In models gbl-$m10$ and gbl-$m10+$ we
excite a specific Fourier mode using a plane wave, which takes the
form of Eq~(\ref{eqn:vz_pert}), as described in
\S~\ref{sec:excite}. These models use the same wavenumbers as model
lin-$m10$ so as to allow a direct comparison of magnetic field
growth. The third model, gbl-rand, uses random pressure and poloidal
velocity perturbations to initially seed the disk disturbance. All of
the global models start with a purely toroidal magnetic field with
$\beta=20$. Models gbl-$m10+$ and gbl-rand are computed on grids with
lower poloidal resolution (roughly 2/3 that of model gbl-$m10$), but
with a factor of three better azimuthal resolution. In the following
section we present some properties of our model disks and demonstrate
that higher $\phi$ resolution to be a crucial ingredient in producing
a sustained, high valued turbulent stress, $<\alpha_{\rm P}>$.

\subsection{Model evolution}
\label{subsec:fiducial}

We begin with a description of the evolution of models gbl-$m10$ and
gbl-$m10+$ (Tables~\ref{tab:alg_test} and
~\ref{tab:global_models}). In this model we adopt an azimuthal
wavenumber which varies with cylindrical radius, $m=m(R)$. We give the
azimuthal wavenumber a radial dependence of $m(R) =m_{\rm crit}(R)/6$,
where the critical\footnote{Defined as the value of $m$ for which
  disturbances grow most rapidly, which follows from equation (2.30)
  of \cite{BH92}. The radial dependence of Eq~(\ref{eqn:mcrit}) stems
  from the \cite{Paczynsky:1980} potential (Eq~(\ref{eqn:pwpot})).}
azimuthal wavenumber for the linear, non-axisymmetric MRI,
\begin{equation}
  m_{\rm crit} = \frac{\sqrt{R}}{R-2} \sqrt{\frac{\beta}{2 T}}. \label{eqn:mcrit}
\end{equation} 
\noindent At $R=20$ in model gbl-$m10$, $m_{\rm crit} = 60$; adopting
$m=m_{\rm crit}/6$ ensures that the corresponding $k_{\rm z}$ and
$k_{\rm R}$ are well resolved by the numerical grid. \cite{BH92} noted
that the fastest growing non-axisymmetric modes occur for $k_{\rm
  z}=m^2/R$, and we also use this relation to calculate $k_{\rm
  z}$. Given that our grid resolution is coarser in $r$ than it is in
$\theta$, we set $k_{\rm R}= k_{\rm z}/2$. 

The initial poloidal velocity perturbations seed the growth of
magnetic field perturbations via the MRI and after roughly $1-2~P^{\rm
  orb}_{30}$ turbulent motions become apparent in the disk body. As
the poloidal magnetic field becomes established throughout the disk
the resulting Maxwell stresses disrupt the disk equilibrium. The
evolution of models gbl-$m10$ and gbl-$m10+$ is largely similar during
the first few orbits of the simulations. Examining the normalized
stress, $<\alpha_{\rm P}>$, shows that there is an initial transient
phase which peaks after a simulation time of roughly $4~P^{\rm
  orb}_{30}$ (Fig.~\ref{fig:global_stress}). Following this,
$<\alpha_{\rm P}>$ gradually decreases until a steady-state is reached
after roughly $12~P^{\rm orb}_{30}$ and the time-averaged stress for
the remainder of the simulation, $\overline{<\alpha_{\rm P}>}= 0.017$
for gbl-$m10$. The time-averaged ratio of the Maxwell stress to the
magnetic energy, $\overline{<\alpha_{\rm M}>}=0.31$, which is below
the values of roughly 0.4 quoted by, for example, \cite{Hawley:2011}
for well resolved turbulence. To investigate the dependence of these
values on the azimuthal resolution of the simulation we have also run
model gbl-$m10+$, which has 12.5 cells/$H$ in the $\phi$ direction
(and a lower resolution in the poloidal direction - see
Table~\ref{tab:alg_test}). The higher $\phi$ resolution clearly
influences the turbulent stresses in the simulation and for model
gbl-$m10$ we find $\overline{<\alpha_{\rm M}>}=0.41$ and
$\overline{<\alpha_{\rm P}>}=0.034$, in agreement with high resolution
shearing-box simulations. The resolvability of the fastest growing MRI
modes (see Eq~(\ref{eqn:resolvability}) and
Fig.~\ref{fig:resolvability}) also clearly show that a higher
azimuthal resolution helps to maintain (or even strengthen) the
poloidal magnetic field - models gbl-$m10$ and gbl-$m10+$ initially
show similar values of $N_{\rm z}$ but largely different values of
$N_{\phi}$, and combined with the evidence mentioned above is evident
that azimuthal resolution is very important for maintaining a healthy
turbulent state \citep[see also the discussion
in][]{Fromang:2006,Flock:2011,Hawley:2011}. In \S~\ref{sec:discussion}
we compare further quantitative measures of the steady-state
turbulence to previous works.

The poloidal magnetic field develops in flux tubes with small spatial
scale, which dissipate magnetic energy via reconnection, heating the
disk.  In Fig.~\ref{fig:H_R} we show the density-weighted and
volume-averaged scaleheight of the disk, $<H/R>$ as a function of
time. In model gbl-$m10$ the scaleheight of the disk increases
initially until $t\sim 8~P^{\rm orb}_{30}$, after which it steadily
declines. This shows that during the initial disk evolution,
dissipation heats the disk more rapidly than the cooling function,
$\Lambda$ (see Eq~(\ref{eqn:cooling_function})) can drive the
temperature back to its initial value. In other words, the dissipative
timescale is shorter than an orbital period. In contrast, for model
gbl-$m10+$, $<H/R>$ remains roughly constant after the initial rise,
which shows that the higher $<\alpha_{\rm P}>$ in this model
(Fig.~\ref{fig:global_stress}) is causing more heating, and a
marginally thicker disk.

\begin{figure}
  \begin{center}
    \begin{tabular}{c}
 \resizebox{85mm}{!}{\includegraphics{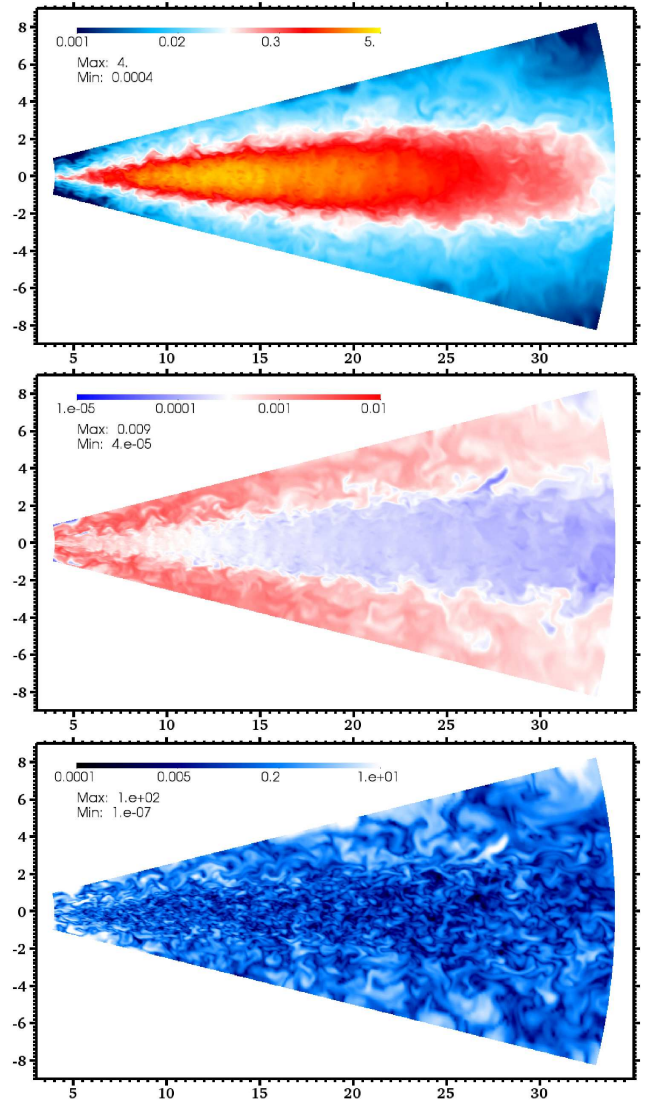}} \\
  \end{tabular}
  \caption{Slices in the poloidal plane from model gbl-$m10+$ at
    $t=14\;P^{\rm orb}_{30}$ showing $\rho$ (upper), $T$ (middle), and
    $\beta^{-1}$ (lower).}
    \label{fig:gbl_beta20}
  \end{center}
\end{figure}

\begin{figure}
  \begin{center}
    \begin{tabular}{c}
\resizebox{80mm}{!}{\includegraphics{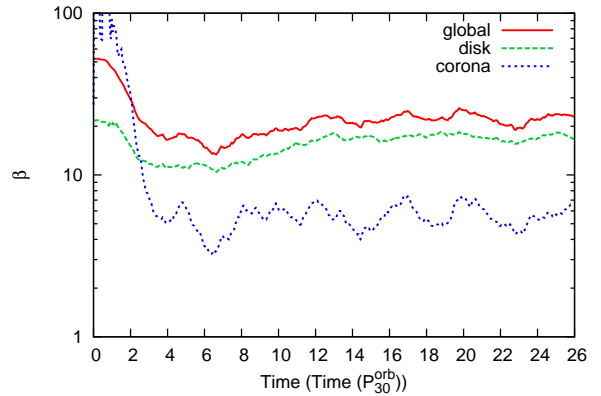}} \\
    \end{tabular}
    \caption{Disk body, corona, and global volume averages of the
      plasma-$\beta$ for model gbl-$m10+$.}
    \label{fig:beta}
  \end{center}
\end{figure}

Fig.~\ref{fig:gbl_beta20} shows a poloidal slice through the disk in
model gbl-$m10+$ at $t=14~P^{\rm orb}_{30}$. During the turbulent
steady state the disk is characterised by a dense, cold, sub-thermally
magnetized core close to the mid-plane and a tenuous, hot,
trans-to-super thermal magnetic field at $z\gtsimm 2H$ (the
corona). Turbulent motions are clearly evident in the plot of
$\beta^{-1}$ in Fig.~\ref{fig:gbl_beta20} with the dominant eddies
appearing to have a larger size in the corona compared to the disk
body. As noted by \cite{Fromang:2006}, such behaviour arises due to
conservation of angular momentum in eddie motions - or wave action -
as small scale eddies rise out of the dense disk mid-plane into the
less dense coronal region. Our intended purpose for the explicit
cooling function, $\Lambda$ (see \S~\ref{subsec:hydromodel} for
details) becomes more apparent from the temperature plot - we aim to
take a step beyond the purely isothermal approximation and towards the
observationally supported picture of a hot corona and cooler disk
body. In Fig.~\ref{fig:beta} we show the volume-averaged
plasma-$\beta$. In the disk body, we find $\overline{<\beta>}=17$ and
for the corona $\overline{<\beta>}=6$. The coronal value is higher
than values of close to one found in previous isothermal
\citep{Miller:2000, Flock:2011} and quasi-isothermal simulations
\citep{Fromang:2006, Beckwith:2011}, which may be attributable to the
lack of any explicit cooling in the corona in our
simulations. However, although the gas in the corona is heated by
dissipation, it does not continually heat up through the simulation,
and in fact remains quasi-steady through the latter half of the
simulation. This contrasts with adiabatic shearing-box simulations
with imposed periodic boundary conditions, in which the gas does heat
up \citep[e.g.][]{Stone:1996, Sano:2004} and demonstrates that when
coronal gas is allowed to freely expand, adiabatic cooling can, to
some extent, balance heating via turbulent dissipation.

\begin{figure}
  \begin{center}
    \begin{tabular}{c}
\resizebox{75mm}{!}{\includegraphics{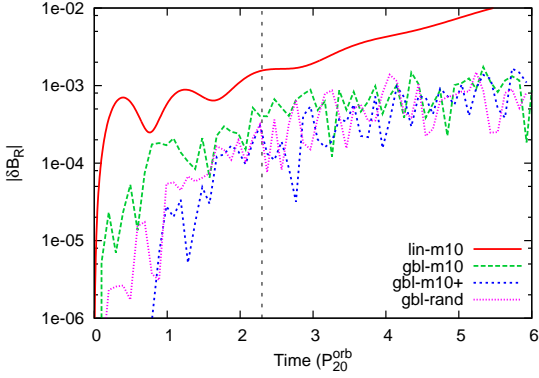}} \\
    \end{tabular}
    \caption{Comparison of $|\delta B_{\rm R} (m=10,k_{\rm z}=5,k_{\rm
        R}=2.5)|$ from models lin-$m10$ (local linear MRI) and models
      gbl-$m10$, gbl-$m10+$, and gbl-rand (fully non-linear global
      simulations). }
    \label{fig:Bcomp}
  \end{center}
\end{figure}

\begin{figure}
  \begin{center}
    \begin{tabular}{c}
\resizebox{75mm}{!}{\includegraphics{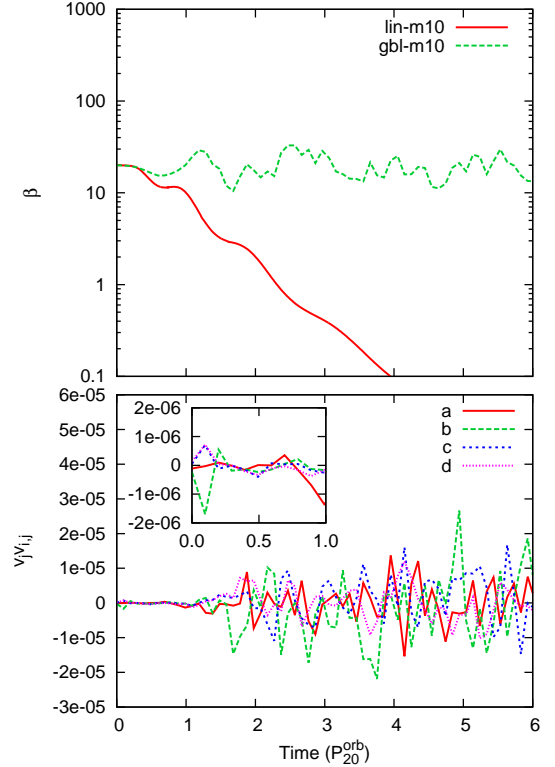}} \\
   \end{tabular}
   \caption{Plasma-$\beta$ from models lin-$m10$ and gbl-$m10$
     (upper), and time evolution of some sample non-linear terms in
     model gbl-$m10$ (lower).}
    \label{fig:comp_fail}
  \end{center}
\end{figure}

\subsection{Comparison with linear MRI growth estimates}
\label{subsec:comp}

In Fig.~\ref{fig:Bcomp} we compare the evolution of $|\delta B_{\rm
  R}|$ for the $m=10$, $k_{\rm z}=5$, $k_{\rm R}=2.5$ mode (measured
in Fourier space - see \S~\ref{subsec:fourier}) for model lin-$m10$
(which describes linear MRI growth - see Tables~\ref{tab:alg_test} and
\ref{tab:local_models}) and models gbl-$m10$ and gbl-$m10+$ (global
simulations which allow fully non-linear evolution - see
Table~\ref{tab:global_models}). Quantifying the initial growth by
deriving approximate growth rates\footnote{An approximate exponential
  growth rate is determined by fitting $|\delta B_{\rm R}| = |\delta
  B_{\rm R}|_{0} \exp(\omega_{\rm approx} t)$.}, $\omega_{\rm approx}$
for the curves shown in Fig.~\ref{fig:Bcomp} , we find that the linear
MRI estimate is matched best by model gbl-$m10+$, with gbl-$m10$ (and
gbl-rand) producing higher growth rates. The higher amplitude
perturbation for model gbl-$m10$ compared to gbl-$m10+$ originates
from the larger amplitude initial velocity perturbation of $0.1~c_{\rm
  s}$ (compared to $0.001~c_{\rm s}$ for gbl-$m10+$ and gbl-rand - see
Table~\ref{tab:global_models}). The agreement between the global
simulations and linear growth estimate (model lin-$m10$) begins to
falter after roughly $2~P^{\rm orb}_{20}$ and growth in $|\delta
B_{\rm R}|$ for the global simulations levels off. From the linear MRI
calculations shown in Fig.~\ref{fig:local_limits} one may anticipate
that the growth in $|\delta B|$ is halted by the magnetic pressure
evolving to equipartition with the gas pressure - which is illustrated
by the vertical dashed line in Fig.~\ref{fig:Bcomp} - and would mean
that one cannot rely on model lin-$m10$ as a predictor for model
gbl-$m10$. However, Fig.~\ref{fig:comp_fail} shows that $\beta$
remains roughly constant in model gbl-$m10$ over the first few
orbits. What then causes the local MRI estimates and the global
simulations to diverge?  In the lower panel of
Fig.~\ref{fig:comp_fail} we plot the evolution of the following
non-linear terms derived from the momentum equations,
\begin{eqnarray}
  a = v_{\rm r} \frac{\partial v_{\rm r}}{\partial r}, b =
  \frac{v_{\theta}}{r} \frac{\partial v_{\rm r}}{\partial \theta},
  \nonumber \\
  c = v_{\rm r} \frac{\partial v_{\theta}}{\partial r}, d =
  \frac{v_{\theta}}{r} \frac{\partial v_{\theta}}{\partial
    \theta}. \nonumber 
\end{eqnarray}
\noindent 
Small spikes in these non-linear terms occur after
$\sim0.1$~orbit. However, after 1-2 orbits considerably larger
fluctuations become apparent, particularly in $b$ which also appears
to have the highest amplitude oscillations of the plotted terms
thereafter. Considering that the linear MRI is seeded by velocity
perturbations through the induction equation (Eqs~(\ref{eqn:dBRdt})
and (\ref{eqn:dBzdt})), the correlation in time between the non-linear
velocity terms becoming active and the growth in $|\delta B_{\rm R}|$
departing from the linear MRI growth predictions is highly suggestive
of non-linear motions causing saturation in the growth of a specific
MRI mode. Furthermore, the small amplitude kicks from these non-linear
terms after 0.1 orbits may explain the early divergence between the
$\beta$ values predicted from model lin-$m10$ and those found from
gbl-$m10$. In this sense the non-linear motions provide saturation to
the initial phase of {\it local} $\delta {\bf B}$ growth. Whether the
non-linear motions are attributable to secondary instabilities feeding
off the linear MRI growth locally \citep[e.g.][]{Goodman:1994,
  Pessah:2007, Pessah:2009, Pessah:2010}, or are due to the onset of
turbulence \citep{Latter:2009} propagating radially outwards through
the disk is unclear and would require an analysis of the non-linear
growth of the non-axisymmetric MRI, which we do not pursue here.

\begin{figure}
  \begin{center}
    \begin{tabular}{c}
\resizebox{80mm}{!}{\includegraphics{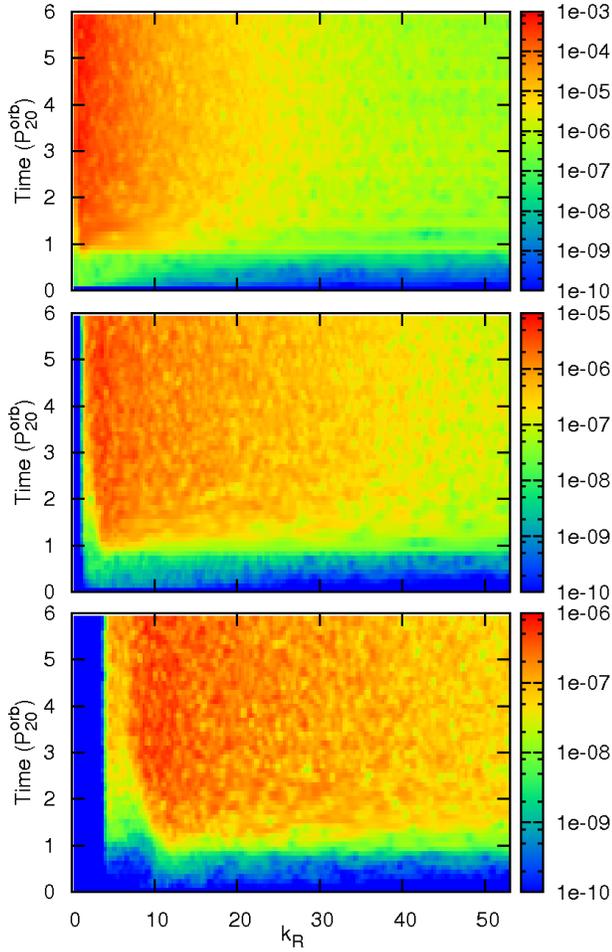}} \\
    \end{tabular}
    \caption{Logarithmic false-colour image showing the time evolution
      of $|\delta B_{\rm R} ({\bf k})|$ in model gbl-$m10+$. Values of
      $\delta B_{\rm R}$ in the disk body were Fourier transformed and
      results are shown at specific values of $m$ and $k_{\rm z}$, and
      the full range of $k_{\rm R }$. From top to bottom: $m=10$ and
      $k_{\rm z}=5$ (low wavenumber), $m=40$ and $k_{\rm z}=80$
      (moderate wavenumber), and $m=120$ and $k_{\rm z}=150$ (high
      wavenumber).}
    \label{fig:kr2d_gbl-m10+}
  \end{center}
\end{figure}

\begin{figure}
  \begin{center}
    \begin{tabular}{c}
\resizebox{80mm}{!}{\includegraphics{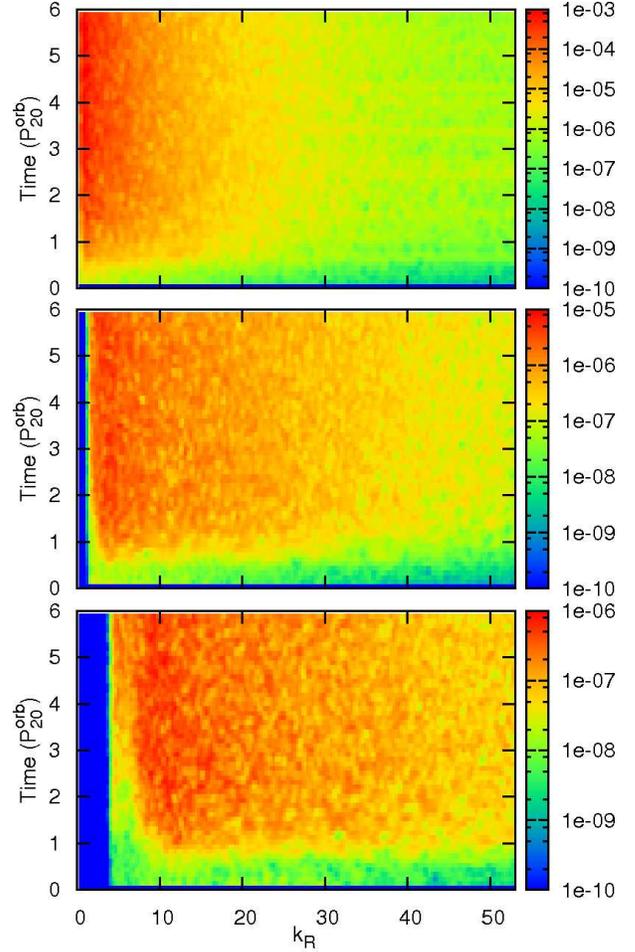}} \\
    \end{tabular}
    \caption{Same as Fig.~\ref{fig:kr2d_gbl-m10+} except for model
      gbl-rand.}
    \label{fig:kr2d_gbl-rand}
  \end{center}
\end{figure}

In summary, comparisons between linear growth calculations and global
simulations highlights a number of potential saturation
mechanisms. Such as, growth of magnetic field perturbations beyond the
weak field limit, and/or growth of the radial wavenumber beyond the
finite limit of the simulation resolution. However, for the
simulations performed in this work, we find that saturation of growth
in magnetic field perturbations correlates well with the onset of
non-linear motions.
\pagebreak
\subsection{Trigger dependence}
\label{subsec:trigger}

A major focus of magnetized accretion disk simulations is to study
properties of the quasi-steady-state turbulence. A necessary test is
whether the turbulent steady state depends on the MRI mode initially
excited, and also whether prohibitive transient behaviour arises due
to the choice of exciting a specific MRI mode. For this purpose we
have computed model gbl-rand which uses the same initial disk as model
gbl-$m10+$ with the difference that the disk is perturbed with random
perturbations in the both the poloidal velocity (amplitude $\delta
v_{0}=0.001~c_{\rm s}$) and gas pressure ($10\%$ amplitude) which
excite a range of MRI modes. Simulation resolution and time-averaged
measures of the turbulent state are listed in
Tables~\ref{tab:alg_test} and \ref{tab:global_models}, respectively.

The evolution of model gbl-rand is very similar to that of model
gbl-$m10+$; the initial perturbations excite the MRI and lead to
growth of $\delta B_{\rm R}$. Both models show similar growth in the
$m=10$, $k_{\rm z}$, $k_{\rm R}=2.5$ mode (Fig.~\ref{fig:Bcomp}) which
one would expect given that this mode is excited with the same
amplitude perturbation. After $3~P^{\rm orb}_{20}$ ($\simeq0.5~P^{\rm
  orb}_{30}$) values of $\delta B_{\rm R}$ become almost identical
between the models irrespective of the differing initial
perturbations. We illustrate this in Figs.~\ref{fig:kr2d_gbl-m10+} and
\ref{fig:kr2d_gbl-rand} in which we show the evolution of $\delta
B_{\rm R}$ in Fourier space for a range of $k_{\rm R}$ values. The
different panels in the figures correspond to low, moderate, and high
wavenumber values for $m$ and $k_{\rm z}$ (relative to the size of the
disk and the Nyquist limit). As mentioned above, $\delta B_{\rm R}$
values are very similar between the two models at $t>3~P^{\rm
  orb}_{20}$. Furthermore, even though we excite a specific low
wavenumber mode in model gbl-$m10$, a wide range of modes rapidly
emerge. We attribute this behaviour to wave-wave coupling and the
onset of a turbulent cascade.

Exciting larger wavenumbers should provide a larger {\it initial} MRI
growth rate (see \S~\ref{sec:excite}), but how does this affect the
evolution of magnetic field perturbations in the global simulations?
In particular, does the wavenumber of the excited MRI mode affect the
globally-averaged saturation stress? In model gbl-rand a white noise
spectrum of perturbations has been excited. Therefore higher
wavenumber modes can contribute to the initial growth phase in
$<\alpha_{\rm P}>$. There is an indication of this from
Fig.~\ref{fig:kr2d_gbl-rand}) where the growth of $|\delta B_{\rm R}|$
at a range of wavenumbers means that the Maxwell stress, and
consequently $<\alpha_{\rm P}>$ will also grow across a range of
wavenumbers. Fig.~\ref{fig:global_stress} shows that $<\alpha_{\rm
  P}>$ does grow faster for gbl-rand compared to gbl-$m10+$ (which
have identical grid resolution), supporting the notion that the growth
in the globally averaged stress due to an ensemble of unstable modes
is higher than for a single wavenumber mode.

\begin{figure}
  \begin{center}
    \begin{tabular}{c}
\resizebox{75mm}{!}{\includegraphics{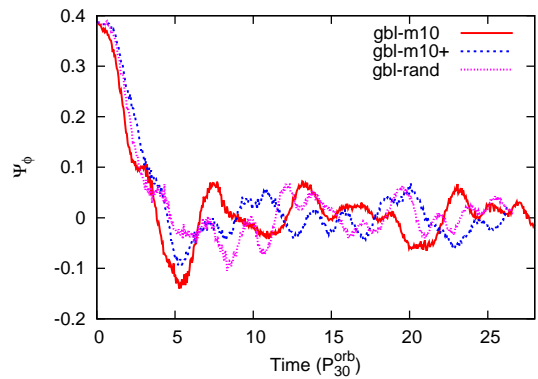}} \\
   \end{tabular}
   \caption{Toroidal magnetic flux at $\phi=0$, $\Psi_{\phi}(\phi =
     0)$ as a function of time for models gbl-$m10$, gbl-$m10+$, and
     gbl-rand. See Eqs~(\ref{eqn:phi_flux}) for the definition of
     $\Psi_{\phi}$.}
    \label{fig:flux}
  \end{center}
\end{figure}

All three models start with a toroidal magnetic field with a net flux,
and during the early evolution of the disk, the combination of
magnetic buoyancy and accretion expels magnetic flux from the disk
body such that by the time the turbulent steady state is reached the
net toroidal magnetic flux of the disk, $\Psi_{\phi}$ is close to
zero. Subsequently, $\Psi_{\phi}$ oscillates about the zero-point with
a period of roughly $5$~orbits (upper panel of Fig.~\ref{fig:flux})
consistent with previous global simulations and suggestive of a dynamo
cycle \citep{Fromang:2006, O'Neill:2011, Beckwith:2011}. All three
models demonstrate this behaviour. However, minor differences in the
toroidal magnetic flux, $\Psi_{\phi}$, are visible between models
gbl-$m10+$ and gbl-rand (Fig.~\ref{fig:flux}). The different models are
slightly out of phase, which is not surprising given the differences
in the transient evolution at early simulation times
(Fig.~\ref{fig:global_stress}). Interestingly, model gbl-rand does not
overshoot when expelling the initial net toroidal flux and thus
settles into dynamo oscillations at a slightly earlier time, which may
explain why the transient phase in $<\alpha_{\rm P}>$ takes a longer
time to fade in this model.

In conclusion, once the disk reaches a turbulent steady-state the disk
retains no knowledge of the MRI mode initially excited. This is
supported by the almost identical time-averaged properties of the disk
noted in Table~\ref{tab:global_models} for models perturbed by a
single low wavenumber mode or an ensemble of modes.

\subsection{Algorithm and resolution dependence}
\label{subsec:alg_test}

In this section we examine the ability of different numerical
algorithms to recover the growth of magnetic field perturbations
resulting from the non-axisymmetric MRI. Comparisons between numerical
simulations and analytical estimates for the {\it axisymmetric} MRI
have been presented by \cite{Hawley:1991} and
\cite{Flock:2010}. Considering that MHD turbulence in accretion disks
produces a predominantly toroidal field it is important to examine how
well numerical algorithms can recover the growth of $\delta {\bf B}$
as a result of the {\it non-axisymmetric} MRI. The setups used are
listed in Table~\ref{tab:alg_test}. The different combinations are
intended to test different orders of reconstruction, parabolic
limiters, and grid resolution\footnote{Our aim is to examine the
  ability of different algorithms to capture
  waveforms. \cite{Balsara:2010} found that Riemann solvers that do
  not resolve the Alfv{\' e}n wave are more likely to lead to
  turbulence dying out as a result of a higher level of numerical
  dissipation. Therefore, we have not endeavoured to test different
  Riemann solvers and we use the HLLD solver of \cite{Miyoshi:2005} in
  all global simulations.}. Reconstruction refers to the order of
accuracy used to interpolate cell interface values (which are then
used in the Riemann solver to calculate fluxes of conserved
variables). Parabolic limiters are used to preserve monotonicity and
prevent extrema from being introduced by the reconstruction step. We
examine the original limiter for PPM proposed by \cite{Colella:1984},
the extremum preserving limiters presented by \cite{Colella:2008}, and
limiters based on reconstruction via characteristic variables
\citep[e.g.][]{Rider:2007}. The aforementioned slope limiters are
respectively denoted ``CW84'', ``CS08'', and ``Char'' in
Table~\ref{tab:alg_test}. The last parameter we vary is the grid
resolution, as this places a constraint on the maximum resolvable
wavenumber, and for this purpose we compute models gbl-$m10$-hr,
gbl-$m10$, gbl-$m10$-lr, gbl-$m10$-llr, and gbl-$m10$-lllr (which have
decreasing resolution). Note that with the exception of models
gbl-$m10+$ and gbl-rand, all models have the same cell aspect ratio
and the same ratio of cells in the disk body to cells in the corona as
gbl-$m10$.

As in \S~\ref{subsec:comp}, we calculate approximate growth rates of
the magnetic field perturbation, $|\delta B_{\rm R}({\bf k})|$ for the
$m=10$, $k_{\rm z}=5$, $k_{\rm R}=2.5$ mode via a Fourier analysis of
the initial simulation evolution. The results are shown in
Table~\ref{tab:alg_test}, which we summarise as follows:
\begin{itemize}
\item The best agreement with the linear MRI growth rate comes from
  model gbl-$m10+$, showing that azimuthal resolution is important for
  properly capturing MRI growth.
\item Within errors the choice of limiter does not make a considerable
  difference to the resulting growth rate.
\item Linear reconstruction produces a comparable growth rate to
  parabolic reconstruction. 
\item There is a consistent trend that growth rates increase with
  increasing resolution (see models gbl-$m10$-lllr, gbl-$m10$-llr,
  gbl-$m10$-lr, gbl-$m10$, and gbl-$m10$-hr). With 4 cells/$H$ in the
  $\phi$ direction the growth rates are converged (within errors) for
  roughly 23-35 cells/$H$ in the $z$ direction. This is a slightly
  lower threshold than the $\gtsimm 40$ cells/$H$ (in the vertical
  direction) found to achieve convergence in the time-averaged
  turbulent stress in stratified shearing-box simulations
  \citep{Davis:2010}. Comparison with model gbl-$m10+$ (which has 12.5
  cells/$H$ in the $\phi$ direction) suggests that convergence in
  global disks may be achieved at lower resolutions when the cell
  aspect ratio is closer to unity.
\end{itemize}

\section{Discussion}
\label{sec:discussion}

With a growing number of studies using stratified shearing boxes with
high resolution and/or a large spatial extent \citep{Shi:2010,
  Davis:2010, Guan:2011, Simon:2011, Simon:2012} and higher resolution
global models \citep{Fromang:2006, Sorathia:2010, Sorathia:2012,
  Flock:2011, Flock:2012, Beckwith:2011,Hawley:2011,Mignone:2012},
quantifying the steady state turbulence and making direct comparisons
between simulations provides a check of both consistency and
convergence.

One of the most popular measures of the steady state is
$\overline{<\alpha_{\rm P}>}$.  In this regard, models gbl-$m10+$ and
gbl-rand produce values of $\sim 0.034$ which is higher than recently
reported by \cite{Beckwith:2011} and, as noted by these authors,
higher than previous global models and a number of high resolution
shearing-box simulations \citep[see][and references
there-in]{Hawley:2011}. We attribute the larger
$\overline{<\alpha_{\rm P}>}$ in our models to a higher azimuthal
resolution than used by \citep{Beckwith:2011}, but also note the
possible indication that higher $\overline{<\alpha_{\rm P}>}$'s may be
more readily achievable in global simulations. Our average
$\overline{<\alpha_{\rm M}>} \sim 0.41$ (for models gbl-$m10+$ and
gbl-rand) is in good agreement with the $\sim 0.36 - 0.4$ achieved by
the highest resolution shearing-box simulations to-date
\citep{Davis:2010, Simon:2011,Simon:2012}. Considering that our models
have a lower number of cells/$H$ than the aforementioned shearing-box
models, there may also be an indication that convergence may be
achieved at lower grid resolutions than in localized models,
potentially due to averaging over a larger volume, and capturing lower
wavenumber eddies.

Comparing models gbl-$m10$ and gbl-$m10+$, strong evidence points to
the grid cell aspect ratio and, in particular, the resolution in the
$\phi$ direction as an important parameter in achieving a high
$\overline{<\alpha_{\rm P}>}$ and $\overline{<\alpha_{\rm M}>}$ (see
the discussion in \citeauthor{Fromang:2006}~\citeyear{Fromang:2006},
\citeauthor{Flock:2011}~\citeyear{Flock:2011},
\citeauthor{Hawley:2011}~\citeyear{Hawley:2011} and
\citeauthor{Sorathia:2012}~\citeyear{Sorathia:2012}). A possible
explanation for this is that the dynamo cycle - which helps to sustain
the turbulent state and involves the MRI as a driving agent - can
operate more effectively at higher frequencies when the cell aspect
ratio is closer to unity.

Related to $\overline{<\alpha_{\rm M}>}$ is the tilt angle,
$\Theta_{\rm tilt}$, where $\sin 2 \Theta_{\rm tilt} = <\alpha_{\rm
  M}>$ \citep{Guan:2009, Beckwith:2011}. It has been argued by
\cite{Sorathia:2012} that this parameter provides a better measure of
convergence than $\overline{<\alpha_{\rm P}>}$, at least in the case
of unstratified turbulence for which the question of convergence in
the absence of explicit dissipation was raised by
\cite{Fromang:2007a}. We find $\Theta_{\rm tilt}\sim9^{\circ}$ for
model gbl-$m10$ which is consistent with previous findings for
stratified global disks \citep{Beckwith:2011, Hawley:2011,
  Flock:2012}. However, models gbl-$m10+$ and gbl-rand have
$\Theta_{\rm tilt}\sim12^{\circ}$ which is comparable to values of
$\sim 11^{\circ}-13^{\circ}$ for shearing box simulations \citep[both
unstratified and stratified, e.g.,][]{Guan:2009, Simon:2012} and also
for recent stratified global disks calculations performed with an
orbital advection algorithm by \cite{Mignone:2012}. These results are
encouraging as they show that global simulations are reaching
sufficient grid resolution to reproduce shearing-box results.

The ratio of directional magnetic energy also provides insight into
convergence and correspondence between simulations. We find,
$\overline{<B_{\rm R}^2>/<B_{\phi}^2>} \sim 0.13$, $\overline{<B_{\rm
    z}^2>/<B_{\rm R}^2>} \sim 0.30$ and $\overline{<B_{\rm
    z}^2>/<B_{\phi}^2>} \sim 0.036$ for models gbl-$m10+$ and
gbl-rand. These values are higher than obtained by \cite{Hawley:2011}
for their global disk simulations, and in some cases only slightly
lower than values found from high resolution shearing-box simulations
\citep{Shi:2010, Davis:2010, Simon:2011, Guan:2011, Simon:2012}.

Interestingly, model gbl-$m10$ produces a sustained stress, albeit
with a lower value than model gbl-$m10+$, but with only 4 cells/$H$ in
the $\phi$-direction. \cite{Flock:2011} found that at least 8
cells/$H$ were required to produce a sustained turbulent stress
\citep[see also][]{Fromang:2006}. However, these authors used linear
reconstruction, whereas we have used parabolic reconstruction which
may permit a sustained stress at a slightly lower resolution.

Finally, we note that we do not see any prominent evidence of
recurring transient phenomena due to linear growth revivals in the
mean magnetic fields, as recently reported by \cite{Flock:2012}. This
may be due to differences in the numerical setup between our models
and those of \citeauthor{Flock:2012}, or perhaps this phenomena occurs
at later times that we have not reached with the simulation runtimes
of our models.

\section{Conclusions}
\label{sec:conclusions}

We have performed global 3D MHD simulations of turbulent accretion
disks which start from fully equilibrium MHD initial conditions. The
local linear theory of the MRI is used as a predictor of the growth of
magnetic field perturbations in the global simulations. Additional
tests have also been performed to compare results obtained from global
simulations using a number of different numerical algorithms and
resolutions to the linear growth estimates. Our main findings are:
\begin{enumerate}[i)]
\item The growth of magnetic field perturbations in the global
  simulations shows good agreement with the linear MRI growth
  estimates during approximately the first orbit of the
  disk. Subsequently, the overwhelming influence of non-linear
  motions, which may be due either to the onset of turbulence or to
  secondary instabilities, saturates the {\it local} growth.
\item The saturated state is found to be independent of the initially
  excited MRI mode, showing that once the disk has expelled the
  initial net flux field and settled into quasi-periodic oscillations
  in the toroidal magnetic flux, the dynamo cycle regulates the global
  saturation stress level. Furthermore, time-averaged measures of
  quasi-steady turbulence are found to be in agreement with previous
  work. In particular, the time averaged stress,
  $\overline{<\alpha_{\rm P}>}\sim0.034$.
\item We find $\overline{<B_{\rm R}^2>/<B_{\phi}^2>} \sim 0.13$ for
  global stratified simulations with 12.5 cells/$H$ in the $\phi$
  direction, which is in good agreement with value found from high
  resolution, stratified, shearing-box simulations. Higher $\phi$
  resolution in the simulation (at least $> 4$ cells/$H$) is required
  to maintain stronger radial and vertical magnetic field, and
  consequently a larger $\overline{<\alpha_{\rm P}>}$.
\item From the numerical algorithms that we tested, the choice of
  reconstruction order or limiter does not significantly alter the
  resulting linear MRI growth rate. Convergence with resolution (for
  the linear MRI growth tests) is found for resolutions of roughly
  $23-35~$cells per scaleheight (in the vertical direction). However,
  above all, a higher azimuthal resolution contributes to a much
  better agreement with linear growth estimates, supporting the push
  for low cell aspect ratio (close to one) in global accretion disk
  simulations.
\end{enumerate}

\subsection*{Acknowledgements}
We thank the referee for a particularly useful report which helped to
significantly improve the paper. This research was supported under the
Australian Research Council's Discovery Projects funding scheme
(project number DP1096417). E.~R.~P thanks the ARC for funding through
this project. This work was supported by the NCI Facility at the ANU
and by the iVEC facility at the Pawsey Centre, Perth, WA.

%\bibliography{rossbib}{}

\appendix
\section{Equilibrium disk solutions}
\label{sec:appendix}

The initial conditions for our simulations are an equilibrium thin
disk with a purely toriodal magnetic field. In the following we
present some analytic solutions which are of use for numerical
simulations of accretion disks and for disks in other environments,
such as starburst galaxies \citep[e.g.][]{Cooper:2008}. These
solutions incorporate more or less arbitrary radial profiles of
density and temperature and a toroidal magnetic field. The latter
involves either a constant ratio of gas-to-magnetic pressure, $\beta$,
radially dependent $\beta$, constant $B_{\phi}$, and variants with
net/zero toroidal magnetic flux.

In axisymmetric cylindrical coordinates ($R,\phi,z$), in steady-state,
and with $v_{\rm R} = v_{\rm z} = B_{\rm R} = B_{\rm z} = 0$, the
induction equation is identically satisfied and we are left with the
two momentum equations:
\begin{eqnarray}
\frac{\partial P}{\partial R} +  \rho \frac{\partial\Phi}{\partial R}
- \frac{\rho v_{\phi}^2}{R} + \frac{1}{2
  R^2} \frac{\partial (R^2 B_{\phi}^2)}{\partial R} &  =  &
0, \label{eqn:rmom_app} \\ 
\frac{\partial P}{\partial z} + \rho \frac{\partial\Phi}{\partial z} +
\frac{1}{2}\frac{\partial B_{\phi}^2}{\partial z} & = &
0, \label{eqn:zmom_app}
\end{eqnarray}
\noindent where the pressure, $P = \rho T$, with $T$ in scaled
units. 

We derive a compatibility condition for the above equations by
subtracting $\partial / \partial R$ of Eq~(\ref{eqn:zmom_app}) from
$\partial / \partial z$ of Eq~(\ref {eqn:rmom_app}), to obtain,
\begin{equation}
\frac{\partial}{\partial z} \left(\frac{v_{\phi}^2}{R}\right) =
\frac{\partial T}{\partial z}\frac{1}{\rho}\frac{\partial
  \rho}{\partial R} - \frac{1}{2 R^2\rho^2}\frac{\partial \rho}{\partial
  z}\frac{\partial (R^2 B_{\phi}^2)}{\partial R} + \frac{1}{\rho
  R}\frac{\partial B_{\phi}^2}{\partial z} - \frac{\partial T}{\partial R}\frac{1}{\rho}\frac{\partial
  \rho}{\partial z} + \frac{1}{2 \rho^2}\frac{\partial \rho}{\partial
  R}\frac{\partial B_{\phi}^2}{\partial z} \label{eqn:comp_cond_app}
\end{equation}
\noindent To solve for the disk equilibrium we take the approach of
using Eq~(\ref{eqn:zmom_app}) to derive an equation for $\rho(R,z)$,
Eq~(\ref{eqn:comp_cond_app}) to acquire $v^2_{\phi}(R,z)$, and
Eq~(\ref{eqn:rmom_app}) to obtain an expression for
$v^2_{\phi}(R,0)$. The resulting equations require boundary conditions
for the run of $\rho$ and $T$ at the disk midplane, which can be
chosen arbitrarily. From here on we take the disk to be isothermal in
height, $T = T(R)$, and firstly consider a disk with a constant,
$\beta = 2P/|B|^2 \equiv 2P/B_{\phi}^2$. Eq~(\ref{eqn:zmom_app}) then
becomes,
\begin{equation}
  \frac{1}{\rho}\frac{\partial \rho}{\partial z} = -\frac{\partial
    \Phi}{\partial z}\left( \frac{ \beta}{1 + \beta}\right)\frac{1}{T(R)} \label{eqn:rho2_app}
\end{equation}
\noindent which integrates to give an expression for the density in
terms of its midplane value,
\begin{equation}
  \rho(R,z) = \rho(R,0) \exp\left(\frac{-\{\Phi(R,z) - \Phi(R,0)\}}{T(R)} \frac{\beta}{1 + \beta} \right). \label{eqn:rho_app}
\end{equation}
\noindent Turning to the rotational velocity, the compatibility
relation (Eq~(\ref{eqn:comp_cond_app})) reduces to,
\begin{equation}
\frac{\partial}{\partial z} \left(\frac{v_{\phi}^2}{R}\right) =
-\left(\frac{1 + \beta}{\beta}\right)\frac{1}{\rho}\frac{\partial
  \rho}{\partial z} \frac{\partial T}{\partial R}
\end{equation}
\noindent which upon integrating and using Eq~(\ref{eqn:rho2_app})
leads to an expression for the azimuthal velocity in terms of its
midplane value,
\begin{equation}
v_{\phi}^2(R,z) = v_{\phi}^2(R,0) + \{\Phi(R,z) -
\Phi(R,0)\}\frac{R}{T}\frac{d T}{d R}.
\end{equation}
\noindent The model is completed with a midplane rotational velocity,
which is determined by substituting Eq~(\ref{eqn:rho_app}) into
Eq~(\ref{eqn:rmom_app}). This gives
\begin{eqnarray}
v_{\phi}^2(R,0) =  R \frac{\partial
  \Phi (R,0)}{\partial R} + \frac{2 T}{\beta} + \left(\frac{1 + \beta}{\beta} \right)\left(\frac{R
    T}{\rho(R,0)}\frac{\partial \rho(R,0)}{\partial R}
 + R \frac{d T}{d R}\right). \label{eqn:vphi0_app}
\end{eqnarray}
\noindent The first term is the square of the Keplerian velocity; the
remaining terms are proportional to the square of the sound speed so
that Eq~(\ref{eqn:vphi0_app}) represents a minor departure from a
Keplerian disk.

A possible variation to the aforementioned disk would be to make
$\beta$ radially dependent, i.e., $\beta=\beta(R)$. For example, one
may choose to make $\beta(R) \propto \sin(k R)$, where $k$ is a radial
wavenumber. In such a case Eq~(\ref{eqn:rho_app}) for $\rho(R,z)$ is
unchanged. However, the expression for the rotational velocity
becomes,
\begin{equation}
  v_{\phi}^2(R,z) = v_{\phi}^2(R,0) + \{\Phi(R,z) -
\Phi(R,0)\}\left(\frac{R}{T}\frac{d T}{d R} - \frac{1}{\beta
    (1+\beta)}\frac{\partial \beta}{\partial R}\right)
\end{equation}
\noindent where, following substitution in Eq~(\ref{eqn:rmom_app}),
\begin{eqnarray}
v_{\phi}^2(R,0) =  R \frac{\partial
  \Phi (R,0)}{\partial R} + \frac{2 T}{\beta} + \left(\frac{1 + \beta}{\beta} \right)\left(\frac{R
    T}{\rho(R,0)}\frac{\partial \rho(R,0)}{\partial R}
 + R \frac{d T}{d R}\right) - \frac{R T}{\beta^2}\frac{\partial
 \beta}{\partial R}. 
\end{eqnarray}

Alternatively, one may desire a disk with a constant
$B_{\phi}(R,z)=B_{\phi 0}$, in which case the magnetic pressure does
not influence the density profile, leading to,
\begin{equation}
  \rho(R,z) = \rho(R,0) \exp\left(\frac{-\{\Phi(R,z) - \Phi(R,0)\}}{T(R)}\right) \label{eqn:rho_const_bphi},
\end{equation}
\noindent and a corresponding velocity profile of,
\begin{equation}
v_{\phi}^2(R,z) = v_{\phi}^2(R,0) + \{\Phi(R,z) -
\Phi(R,0)\}\frac{R}{T}\frac{d T}{d R} + v^2_{\rm A \phi}(R,z) -
v^2_{\rm A \phi}(R,0),
\end{equation}
\noindent with,
\begin{eqnarray}
v_{\phi}^2(R,0) =  R \frac{\partial
  \Phi (R,0)}{\partial R} + v^2_{\rm A \phi}(R,0) + \left(\frac{R
    T}{\rho(R,0)}\frac{\partial \rho(R,0)}{\partial R}
 + R \frac{d T}{d R}\right), 
\end{eqnarray}
and where the Alfv\'{e}n speed, $v_{\rm A \phi}(R,z) =
B_{\phi 0}/\sqrt{\rho(R,z)}$.

As mentioned above, the radial profiles $\rho(R,0)$ and $T(R)$
required to complete the disk model may be chosen arbitrarily, subject
to boundary constraints at the outer disk edge. As an example, we use
simple functions inspired by the \cite{Shakura:1973} disk model,
modified by truncation of the density profile at a specified outer
radius:
\begin{eqnarray}
  \rho(R,0) & = & \rho_{0} f(R, R_{0},R_{\rm out})
  \left(\frac{R}{R_{0}}\right)^{\epsilon}, \label{eqn:density_profile_app}\\
  f(R, R_{0},R_{\rm out}) & = &  \left(\sqrt{\frac{R_{\rm out}}{R}} +
    \sqrt{\frac{R_{\rm 0}}{R}} - \sqrt{\frac{R_{\rm out}R_{0}}{R^2}} - 1
  \right), \label{eqn:fR}\\
  T (R) & = & T_{\rm 0} \left(\frac{R}{R_{0}}\right)^{\chi}, \label{eqn:temp_app}
\end{eqnarray}
\noindent where $\rho_{0}$ sets the density scale, $R_{0}$ and $R_{\rm
  out}$ are the radius of the inner and outer disk edge, respectively,
and $\epsilon$ and $\chi$ set the slope of the density and temperature
profiles, respectively. The tapering function, $f(R, R_{0},R_{\rm
  out})$ is used to truncate the disk at an inner and outer radius. In
practice this function is normalized to give $\rho(R_{\rm max},0) =
\rho_{0}$, where the radius of peak density, $R_{\rm max}$ is given by
the positive root of the quadratic resulting from taking $\partial
/ \partial R$ of Eq~(\ref{eqn:density_profile_app}), namely,
\begin{equation}
  \sqrt{R_{\rm max}} = \frac{a}{2}(\sqrt{R_{\rm out}} + \sqrt{R_{0}}) +
  \frac{1}{2}\sqrt{a^2 (\sqrt{R_{\rm out}} + \sqrt{R_{0}})^2 - 4(1 -
    \epsilon^{-1})\sqrt{R_{0}R_{\rm out}} }
\end{equation}
\noindent where $a=1 - (2\epsilon)^{-1}$. Once $R_{\rm max}$ is known
it is straightforward to renormalize the density profile.

Finally, studies of turbulent dynamos in magnetized disks are often
concerned with the net flux of the magnetic field
\citep[e.g.][]{Brandenburg:1995,Hawley:1996,Fromang:2006}. For the
initially purely toroidal field we have adopted in this paper the net
flux of the disk is given by $\Psi_{\phi} = \int \int B_{\phi} dR
dz$. Noting that in the above derivations we have used $\beta$ to
relate $B_{\phi}^2$ to $P$, meaning $B_{\phi} = \pm \sqrt{2 \rho T /
  \beta}$, i.e. we are free to choose the sign of
$B_{\phi}$. Therefore, if a net flux field is required then one may
set the sign of $B_{\phi}$ the same everywhere, whereas if one desires
a zero-net flux field then, for example, one may choose to make
$B_{\phi}$ anti-symmetric about the disk midplane.

\section{Fourier transform in cylindrical coordinates}
\label{sec:cft}

We wish to evaluate the Fourier transform $F(\bf k)$ of a function $f({\bf r}) = f(R,\phi,z)$ expressed in terms of cylindrical polar coordinates $(R,\phi,z)$. The definition of the Fourier transform is
\begin{equation}
F({\bf k}) = \int \exp (i {\bf k \cdot r}) f({\bf r}) \> d^3x
\end{equation}
Cylindrical coordinates in real and Fourier space are expressed via the following equations:
\begin{equation}
\begin{array}{ c c l c c c l}
x & = & R \cos \phi & \quad & k_{\rm x} & = & k_{\rm R} \cos \psi \\
y & = & R \sin \phi & \quad & k_{\rm y} & = & k_{\rm R} \sin \psi \\
z & = & z           & \quad & k_{\rm z} & = & k_{\rm z}
\end{array}
\end{equation}
Hence,
\begin{equation}
{\bf k \cdot r} = k_{\rm R} R \, \cos(\phi - \psi) + k_z z
\end{equation}
and
\begin{equation}
F({\bf k}) = \int_V \exp [i (k_{\rm R} R \, \cos(\phi - \psi) + k_z z)] \, f(R,\phi,z) \> R dR d\phi dz
\end{equation}
where $V$ is the computational region, usually of the form:
\begin{equation}
R_0 < R < R_1 \qquad 0 < \phi < 2 \pi \qquad -z_0 < z < z_0
\end{equation}

We begin by constructing a Fourier \emph{series} in the periodic azimuthal coordinate $\phi$:
\begin{equation}
f (R, \phi,z) = \sum_{m=-\infty}^\infty f_m(R,z) \, e^{- i m \phi}
\end{equation}
where the coefficients $f_m(R,z)$ are given by:
\begin{equation}
f_m(R,z) = \frac {1}{2\pi} \int_0^{2 \pi} f (R, \phi, z) e^{i m \phi} \> d \phi
\end{equation}
We now make the change of angular variable $\chi = \phi - \psi$; the integration over $\chi$ is still over the interval $[0,2\pi]$ since all of the angular functions within the integrand have period $2 \pi$. The Fourier transform can now be expressed as:
\begin{equation}
F ( {\bf k}) = \sum_{m=-\infty}^\infty e^{i m \psi}
\int_{R_0}^{R_1}
 \left[   \int_{-z_0}^{z_0} e^{i k_z z}f_m(R,z)  
\left[ \int_0^{2\pi}e^{i (- m \chi + k_{\rm R} R \cos \chi)} \> d \chi   \right] \> \> dz \right]  R dR
\end{equation}
The angular integral can be expressed in terms of Bessel functions ($J_m(k_{\rm R} R)$):
\begin{equation}
\int_0^{2\pi}e^{i (- m \chi + k_{\rm R} R \cos \chi)} \> d \chi = 2 \pi \,  i^m J_m(k_{\rm R} R)
\end{equation}
Hence,
\begin{equation}
F({\bf k}) = 2 \pi \sum_{m=-\infty}^\infty i^m \, e^{i m \psi} \>
\int_{R_0}^{R_1} J_m(k_{\rm R} R) \> \left[  \int_{-z_0}^{z_0} e^{i k_z z} f_m(R,z) \> dz \right] \> R \, dR
\label{Fkcyl}
\end{equation}

Equation~(\ref{Fkcyl}) defines the following procedure:\\
\begin{enumerate}
\item Evaluate the angular coefficients:
\begin{equation}
f_m (R,z) = \frac {1}{2\pi} \int_0^{2 \pi} f(R,\phi, z) e^{im \phi} \> d \phi
\end{equation}
\item Then perform the integration in the $z$ direction:
\begin{equation}
F_m(R,k_z) = \int_{-z_0}^{z_0} e^{i k_z z} f_m(R,z) \> dz
\end{equation}
\item Finally, perform the (truncated) Hankel transform in the radial direction:
\begin{equation}
\hat F_m(k_{\rm R}, k_z) = \int_{R_0}^{R_1} R J_m(k_{\rm R} R) F_m(R,k_z) \> R \, dR
\end{equation}
\item The Fourier transform of $f(R,\phi,z)$ is
\begin{equation}
F(k_{\rm R}, \psi, k_z) = 2 \pi \sum_{m=-\infty}^\infty i^m \, e^{i m \psi} \> \hat F_m(k_{\rm R}, k_z) 
\end{equation}
\end{enumerate}

Since the input data for $f(R,\phi,z)$ are on a grid, the azimuthal,
vertical and radial wave numbers, $m, k_{\rm z}$ and $k_{\rm R}$, are
limited by the Nyquist limit.  Let the number of intervals in each
coordinate direction be $(n_{\rm R}, n_\phi, n_{\rm z})$ and the grid
increments be $(\Delta R, \Delta \phi, \Delta z) = [(R_1-R_0)/n_{\rm
  R}, 2 \pi/n_\phi, 2 z_0/n_{\rm z} ]$.  The grid coordinates are
$R_u, \phi_v, z_w$ where:
\begin{equation}
\begin{aligned}
R_u  & =  R_0 + u \, \Delta R  & \quad   u & =  0,~1,...,~n_{\rm R}-1\\
\phi_v & =  v \, \Delta \phi     & \quad   v & =  0,~1,...,~ n_\phi -1\\
z_w  & =  -z_0 + w \, \Delta z & \quad   w & =  0,~1,...,~n_{\rm z} -1
\end{aligned}
\end{equation}

The expressions for the azimuthal $f_m(R,z)$ and vertical $F_m(R,k_z)$
parts of the Fourier transform can be approximated by discrete Fourier
transforms as follows:
\begin{equation}
\begin{aligned}
f_m(R_u,z_w) & \approx  \frac {1}{n_\phi} \sum_{v=0}^{n_\phi-1} f(R_u,\phi_v,z_w) \> \exp [2 \pi i m v /n_\phi] 
& \quad  m & = 0,~1,...,~n_\phi/2 \\
F_m (R_u,k_{\rm z})  & \approx  \frac {2 z_0}{n_{\rm z}} \exp[-i
k_{\rm z} z_0] \> \sum_{w=0}^{n_{\rm z}-1} f_m (R_u,z_w) \> 
\exp [2 \pi l w / n_{\rm z}]  
&\quad   l & = 0,~1,...,~n_{\rm z}/2 
&\quad k_{\rm z} & = \frac {\pi}{z_0} \, l
\end{aligned}
\label{ft_phiz}
\end{equation}
The radial transform can be evaluated from
\begin{equation}
\hat F_m(k_{\rm R}, k_z) = k_{\rm R}^{-2} \int_{k_{\rm R} R_0}^{k_{\rm R} R_1} s J_m(s) F_m(s/k_{\rm R},k_z) \> ds
\end{equation}

More accurate versions of equations~(\ref{ft_phiz}) may be evaluated using the approach given in \citet{Press:1986}.

\label{lastpage}

\end{document}